\documentstyle[preprint,prb,aps,psfig]{revtex}
\begin{document}

\title{Microscopic self-consistent theory of Josephson junctions including 
dynamical electron correlations.}

\author{ P. Miller and J. K. Freericks }

\address{
Department of Physics, Georgetown University, \\
 Washington, D.C. 20057-0995, U.S.A. }
\maketitle
\begin{abstract}
{\sloppy 
We formulate a fully self-consistent, microscopic model to study the 
retardation and correlation effects of the barrier within a Josephson junction. 
The junction is 
described by a series of planes, with electronic correlation included through 
a local self energy for each plane. 
We calculate current-phase relationships for various junctions, which include 
non-magnetic impurities in the barrier region, 
or an interfacial scattering potential. 
Our results indicate that the linear response of the supercurrent to phase 
across the barrier region is a good, but not exact indicator of the critical 
current. Our calculations of the local density of states show the 
current-carrying Andreev 
bound states and their energy evolution with the phase difference across 
the junction. 

We calculate the figure of merit for a Josephson junction, which is the 
product of the critical current, $I_{c}$, 
and the normal state 
resistance, $R_{N}$, for junctions with different barrier materials. 
The normal state resistance is calculated using 
the Kubo formula, for a system with zero current flow and no superconducting 
order. Semiclassical 
calculations would predict that these two quantities are determined by the 
transmission probabilities of electrons in such a way that the product 
is constant for a given superconductor at fixed temperature. 
Our self-consistent solutions for different types of barrier indicate that 
this is not the case. We suggest some forms of barrier which could increase 
the $I_{c}R_{N}$ product, and hence improve the frequency response of a 
Josephson device.
}
\end{abstract}


\section{Introduction}

Some of the most successful and most promising electronics 
applications of superconductors 
involve Josephson junctions, where two superconducting regions are coupled 
through a barrier region, made from a non-superconducting material~\cite{Jose}. 
The drive 
to make electronic devices using Josephson junctions has been motivated by 
their naturally high operating frequencies, far in excess of the speeds 
obtainable in 
standard silicon technology. The operational frequency of an ideal Josephson 
junction using rapid single flux quantum (RSFQ) logic 
is $I_{c} R_{N} e /\hbar $, where $I_{c}$ is the critical current, 
and $R_{N}$ is the normal state resistance of the 
junction~\cite{McCum,McCum2,Likh,Likh2,Likh3}. 
Hence a maximum 
value of the product, $I_{c}R_{N}$ leads to optimal performance. 
At present, low temperature superconductors have been used to produce
junctions with $I_{c}R_{N}$ products up to 1mV, and operational speeds 
reaching $770$ GHz~\cite{Likh3,Patel}, while junctions of 
high-temperature superconductors have achieved $I_{c}R_{N}$ products 
reaching from $1$ mV to $\approx 20$ mV~\cite{Likh4,Char}. 
The choice 
of barrier material strongly affects both $I_{c}$ and $R_{N}$ (typically poor 
conductors increase $R_{N}$ but reduce $I_{c}$) so it is appropriate to study 
in detail what properties of the junction modify $I_{c}$, $R_{N}$, and 
$I_{c}R_{N}$. Indeed, it has been suggested~\cite{Beas}, that the maximum 
value of the product would be reached when the barrier material is close 
to a metal-insulator transition, so that the system is near the cross-over 
from an SNS to an SIS junction. 
In order to investigate such non-trivial barrier materials, we develop a 
microscopic model of a 
Josephson junction, which self-consistently incorporates the 
dynamical correlation effects of the 
electrons in calculating the conductance of the junction and the critical 
current.

Our model is also appropriate for ballistic junctions that have relatively
pure barrier regions so that the extent of the barrier is smaller than its
mean-free-path but larger than its proximity-effect-induced 
superconducting correlation length.  Much work has progressed on these
systems recently, with a concentration on Niobium-Indium Arsenide-Niobium
junctions or Niobium-Silicon-Niobium junctions.  Recent work includes
an examination of subgap structures and current deficits~\cite{chrestin},
quasiparticle reflection effects and spikes in the conductance~\cite{bastian},
effects of multiple Andreev reflection and noise in stacked 
junctions~\cite{thomas}, and an investigation of a tunable junction that
can be altered from an ordinary junction to a $\pi$-junction by driving
the barrier into a nonequilibrium state via a transverse electrical 
current~\cite{klapwick}.  The above-mentioned  experimental and theoretical work
has concentrated on nonequilibrium effects that occur at finite voltages.
This contribution will address only equilibrium and linear-response properties
of the microscopic model for Josephson junctions, but our work can be extended to
examine such nonequilibrium effects as well.

Josephson junctions consist of two superconducting regions coupled through an 
intermediate region which is not naturally superconducting. The main 
characteristics of a Josephson junction can be understood by just 
considering two superconductors coupled together with a 
single, energy independent transmissivity parameter. 
The dc Josephson effect, of a direct supercurrent 
through the junction at zero external voltage, and the ac Josephson effect of 
an oscillating, dissipative current at finite voltage were both predicted from 
such a simple model~\cite{Jose}. 
Further details of the I-V characteristics, such as 
excess current and sub-harmonic gap structure due to Andreev bound states were 
explained~\cite{BTK,KBT}, 
by matching boundary conditions across a step-function in the 
superconducting potential, and including an adjustable scattering 
potential at the interfaces~\cite{Arno,Arno2}. 
Analytic calculations have been performed to 
provide current-phase relationships~\cite{Ambe,Kuli,Furu,Tana}, 
representing the barrier as a 
single scattering potential with no spatial extent. In this contribution we 
compare these traditional approaches with a more microscopic model.
 
Originally the term 
Josephson junction referred to superconductor - insulator - superconductor 
($SIS$) tunnel
junctions, but here we use the term in its common usage to refer to all 
``weak links'', including 
superconductor - normal metal - superconductor ($SNS$) junctions. In 
the latter case, the barrier region can occupy tens to hundreds 
of nanometres, as the 
proximity effect ensures that the superconductivity extends into the 
barrier region, decaying on the scale of the coherence length, 
$\xi_{N} = \hbar v_{F}/k_{B}T$ when in the ballistic regime (where the 
coherence length is longer than the mean free path in the barrier). 

\begin{figure}
  \centerline{\psfig{figure=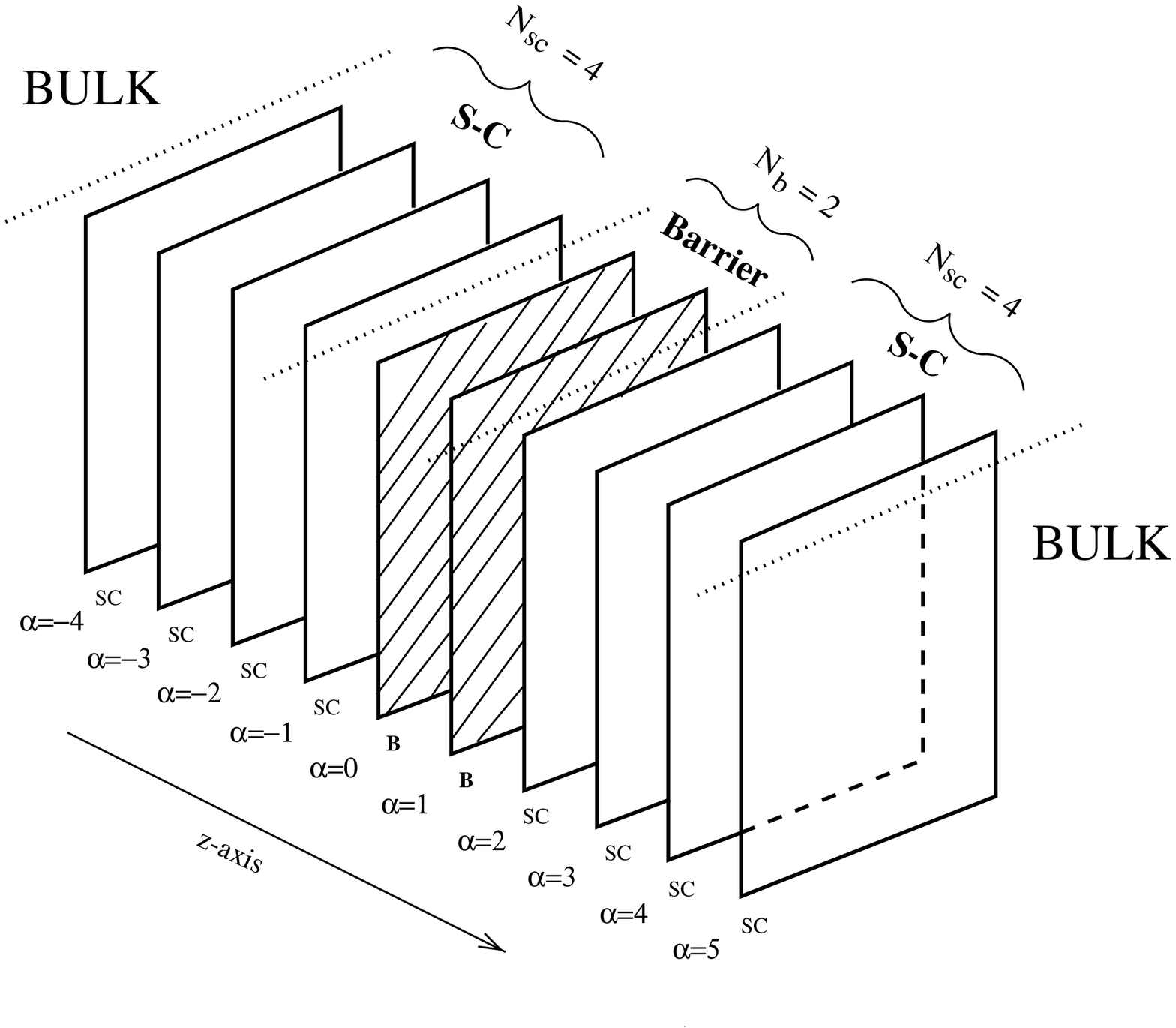,width=7.0in}}
  \caption{Microscopic stacked planar geometry of a Josephson junction. 
The sandwich of $N=10$ planes; $N_{sc}=4$ superconducting planes 
coupled to a bulk superconductor on the left and $N_{b}=2$ barrier 
planes on the right, followed by a further $N_{sc}=4$ superconducting planes 
coupled to another bulk superconductor on the right. 
The system is allowed to have spatial inhomogeneity only within the $N$ 
modeled planes, but the calculations are always for an infinite system.}
\label{fig:Planes}
\end{figure}

In this paper, we present a method for studying the effects of the barrier 
region on the strength of superconductivity it can support, and hence on the 
supercurrent it can maintain. Effects of electron correlations are 
incorporated 
within the dynamic mean field theory~\cite{DMFT1,DMFT2,DMFT3,DMFT4}, 
which leads to a local, 
frequency-dependent self-energy, which we allow to vary from one plane to the 
next, while assuming it to remain constant within individual planes. 
Hence the three-dimensional system becomes inhomogeneous in the one dimension 
where the current flows (which we label the z-direction). We model the system 
with two sets of $N_{SC}$ planes of superconducting material coupled each 
on one 
side to the bulk superconductor, and on the other side sandwiching $N_{b}$ 
planes of barrier material as depicted in Fig.~\ref{fig:Planes}. The total 
number of planes modeled self-consistently is $2N_{SC}+N_{b} = N$. 
We note that 
while in this paper we concentrate on applying the method to Josephson 
junctions, our method would also be applicable to theoretical 
studies of correlated electrons at surfaces, single interfaces, 
or multiple interfaces, as first 
demonstrated by Potthoff and Nolting~\cite{Pott}. 

An important advantage of our scheme is that the different 
(material specific) microscopic models 
that best describe any particular material can be coupled together across 
the planes. Hence a superconductor with electron-phonon coupling 
described by a Holstein model~\cite{Holst} 
(or even using the appropriate  $\alpha^{2}F$ 
in Migdal-Eliahberg theory) can be 
connected to a metallic region with impurities described by the 
Falicov-Kimball model (or a correlated Hubbard model), and so 
on. The local self energy for each plane is calculated independently, 
according to the model best-suited to the particular material. Once a set of 
local self-energies are evaluated for each plane, the Green's functions are 
calculated by finding the inverse of an infinite matrix which includes 
planes of bulk superconductor extending in the positive and negative 
z-direction. The inversion process, which is made tractable by a continued 
fraction representation, couples all of the different planes together, such 
that a change in the self energy on one plane affects the local interacting 
Green's functions on all other planes, 
particularly those nearby in real space. 

In order for a 
supercurrent to flow, a phase gradient must be applied to the superconductor 
and across the barrier region. The critical current, $I_{c}$, is reached, 
when the planes with the lowest superconducting order, typically at the 
center of the barrier region, can no longer support the necessary phase 
gradient to maintain current continuity. 
In our model we find the supercurrent as a function of phase variation across 
the barrier, by solving the system self-consistently at each set of phases. 
The self-consistency is crucial~\cite{Annet,Annet2}, 
as the existence of a current flow affects the 
value of the superconducting order parameter, both inside and outside the 
barrier region. 

Section II contains a detailed description of our method, including the 
physical 
approximations used, and the general computer algorithm. 
In Section III we analyze the current-phase relationships of our 
results in terms of the transmissivity of a barrier. 
The results presented in Section IV demonstrate the efficacy of the 
method to solve some simple models of Josephson junctions, using the 
Bogoliubov-de Gennes equations, where 
we demonstrate the effects of 
self-consistency, in particular on the superconducting order and 
electron density. Section V includes the results for barriers with impurity 
scattering, with 
a description of the calculations of 
normal-state resistance, and results of resistance and $I_{c}R_{N}$ products. 
We conclude in Section VI with some comments on 
the results, and suggestions of other situations well-suited to our model.

\section{Method}

\subsection{Model}

Our method consists of two stages. First, we determine the properties of the 
bulk boundary 
regions for a uniform system. When no current flows in the 
bulk (homogeneous) superconductor, 
dynamical mean field theory~\cite{DMFT1,DMFT2,DMFT3,DMFT4} 
is employed to determine the 
local self energy, using the local approximation. When a uniform current 
is flowing, one must include a uniform phase variation in the superconducting 
order parameter, $\Delta(z)$, for the bulk system. 
Next, we solve the inhomogeneous 
problem self-consistently by iteration, for a number of planes, $N$, 
coupled on either side to the uniform 
bulk solution. That is, giving the plane an index, $\alpha$, 
and defining the 
central link of the junction to be between planes $\alpha=0$ and 
$\alpha=1$, we include 
$N_{b}$ barrier planes, surrounded by $N_{sc}$ self-consistently calculated 
superconductor planes on each side, such that $N=N_{b}+2N_{sc}$ and planes 
with index, $\alpha<1-N/2$ or 
$\alpha>N/2$ are invariant homogeneous bulk planes 
(see Fig.~\ref{fig:Planes}). 
Typically, we solve systems with $N_{sc}=30$, which is significantly greater 
than the coherence length for our bulk superconductor 
($\xi = \hbar v_{F}/\Delta \approx 10a$, where $a$ is the lattice spacing). 
We observe that, except when 
very close to $T_{c}$, the superconducting order has completely healed 
from its disruption at the interface, by the time we reach the 
planes at the bulk superconductor boundary (planes 
with $\alpha \approx \pm N/2$).

We describe the system with the following tight-binding model:
\begin{equation}
\hat{H} = \sum_{i,j,\sigma}  t_{ij} \hat{c}^{\dagger}_{i\sigma} 
\hat{c}_{j\sigma}  + 
\sum_{i} U_{i} \left(\hat{c}^{\dagger}_{i\uparrow} 
\hat{c}_{i\uparrow} - \frac{1}{2} \right) \left( 
\hat{c}^{\dagger}_{i\downarrow} \hat{c}_{i\downarrow} - \frac{1}{2} \right)
+\sum_{i,\sigma}U^{FK}_{i} \hat{c}^{\dagger}_{i\sigma} \hat{c}_{i\sigma}
w_{i}
\label{eq:hhub}
\end{equation}
where $\hat{c}^{\dagger}_{i\sigma}$ and $\hat{c}_{i\sigma}$ are fermionic 
operators which respectively create and destroy an 
electron of spin-$\sigma$ in a single Wannier (tight-binding) state on the
 lattice site $i$;
\[ t_{ij} = \left\{ \begin{array}{cl}
			\epsilon_{\alpha} 	& \mbox{if 
$i=j$ on plane $\alpha$,} \\*
			-t_{\alpha}  & \mbox{if $i$  and 
$j$ are neighboring sites on the same plane, $\alpha$,} \\*
-\sqrt{t_{\alpha}t_{\alpha'}} & \mbox{if $i$ and 
$j$ are neighboring sites on consecutive 
planes, $\alpha$ and $\alpha'$,}\\*
			0		& \mbox{otherwise,}
			\end{array}
		\right. \]
with $-t_{\alpha}$ the overlap or hopping integral for the 
$\alpha$-th plane, 
$\epsilon_{i}$ the local site energy, $U_{i}$ the renormalized on-site, 
Hubbard interaction energy, $U^{FK}_{i}$ the impurity potential, and 
$w_{i}=1$ if there is an impurity on site $i$, and $w_{i}=0$ otherwise. 
The magnitude of the hopping integral in the superconducting 
region, $t$, is constant, and defines our energy scale for the 
entire system ($t=1$). In all results that we present here, the 
superconducting region has an attractive Hubbard interaction, 
$U_{i}=-2$ and no impurities ($w_{i}=0$ for all sites on planes 
$\alpha<1-N_{b}/2$ and $\alpha>N_{b}/2$). 
We utilize the spinless 
Falicov-Kimball model to describe non-magnetic charge 
impurities within the barrier region. The 
interaction between dopant atoms and conduction electrons is attractive, 
and represented by a negative $U^{FK}_{i}$. 
The average impurity concentration within the barrier is given by, 
$\rho_{imp} = \frac{1}{N_{b}}\left(\frac{a}{L}\right)^{2} \sum w_{i}$, 
where $N_{b}\left(\frac{L}{a}\right)^{2}$ is the total number of barrier 
sites. We also consider systems where there are no impurities within the 
barrier, but there is a Hubbard interaction for sites within the barrier, 
$U_{i}=U_{b}$, that differs from $U_{i}=-2$ in the superconductor. In 
addition, we have included systems where the hopping integral, 
$t_{\alpha}$, differs in the barrier region, and where the 
interfacial planes ($\alpha=1-N_{b}/2$ and $\alpha=N_{b}/2$ only) have a 
non-zero local on-site potential, $\epsilon_{\alpha}=V_{Int}$.  

\subsection{Green's Function Calculations}

We use the matrix formulation of Nambu~\cite{Namb} 
for the Green's function, $\underline{\underline{G}}
({\bf r}_{i},{\bf r}_{j},i\omega_{n})$, between two lattice sites 
${\bf r}_{i}$ and ${\bf r}_{j}$ at the Matsubara frequency, 
$i\omega_{n} = i\pi(2n+1)k_{B}T$, 
\begin{equation}
\underline{\underline{G}}({\bf r}_{i},{\bf r}_{j},i\omega_{n}) = 
\left( \begin{array}{cc}
G({\bf r}_{i},{\bf r}_{j},i\omega_{n}) & 
F({\bf r}_{i},{\bf r}_{j},i\omega_{n}) \\*
\overline{F}({\bf r}_{i},{\bf r}_{j},i\omega_{n}) & 
- G^{*}({\bf r}_{i},{\bf r}_{j},i\omega_{n}) 
\end{array} \right) ,
\end{equation}
and the corresponding local self energy, 
\begin{equation}
\underline{\underline{\Sigma}}({\bf r}_{i},i\omega_{n}) = 
\left( \begin{array}{cc}
\Sigma({\bf r}_{i},i\omega_{n}) & \phi({\bf r}_{i},i\omega_{n}) \\*
\phi^{*}({\bf r}_{i},i\omega_{n}) & - \Sigma^{*}({\bf r}_{i},i\omega_{n})) 
\end{array} \right).
\end{equation}
The diagonal and off-diagonal Green's functions are 
defined respectively as:
\begin{eqnarray}
G({\bf r}_{i},{\bf r}_{j},i\omega_{n}) & = & 
-  \int^{\beta}_{0} 
d\tau 
\exp (i\omega_{n}\tau) 
\left<{\rm T}_{\tau} 
\hat{c}_{j\sigma}(\tau) \hat{c}^{\dagger}_{i\sigma}(0) \right>,
\\* 
F({\bf r}_{i},{\bf r}_{j},i\omega_{n}) & 
= & - \int^{\beta}_{0} d\tau 
\exp (i\omega_{n}\tau) 
\left<{\rm T}_{\tau} 
\hat{c}_{j\uparrow}(\tau) \hat{c}_{i\downarrow}(0) \right>,
\end{eqnarray}
where ${\rm T}_{\tau}$ denotes time-ordering in $\tau$ and 
$\beta=1/(k_{B}T)$.

The self energies and Green's functions are coupled together through 
Dyson's equation,
\begin{equation}
\underline{\underline{G}}({\bf r}_{i},{\bf r}_{j},i\omega_{n})
= \underline{\underline{G}}^{(0)}({\bf r}_{i},{\bf r}_{j},i\omega_{n})
+ \sum_{l} 
\underline{\underline{G}}^{(0)}({\bf r}_{i},{\bf r}_{l},i\omega_{n})
\underline{\underline{\Sigma}}({\bf r}_{l},i\omega_{n}) 
\underline{\underline{G}}({\bf r}_{l},{\bf r}_{j},i\omega_{n}),
\label{eq:Dys}
\end{equation}
where we have included the local approximation for the self energy, 
$\underline{\underline{\Sigma}}({\bf r}_{i},{\bf r}_{j},i\omega_{n})
=  \underline{\underline{\Sigma}}({\bf r}_{i},i\omega_{n})\delta_{ij}$, 
[which can be relaxed if we use the Dynamical Cluster Approximation 
(DCA)~\cite{DCA1,DCA2}]. 
The non-interacting Green's function, 
$\underline{\underline{G}}^{(0)}({\bf r}_{i},{\bf r}_{j},i\omega_{n})$ 
is diagonal in Nambu space, with upper diagonal component given by:
\begin{equation}
G^{(0)}({\bf r}_{i},{\bf r}_{j},i\omega_{n}) = \int d^{3}{\bf k} 
\frac{ \mbox{\rm e}^{i {\bf k}\cdot({\bf r}_{i}-{\bf r}_{j}) } }
{ i\omega_{n} -\varepsilon_{{\bf k}} + \mu }.
\end{equation}
We emphasize that $\underline{\underline{G}}^{(0)}$ is the non-interacting 
Green's function and is {\it not} the effective medium of an 
equivalent atomic problem (see below for athe detailed algorithm used to 
solve the dynamical mean field theory). 
A major innovation in our work is to 
utilize an efficient hybrid real-space---momentum-space 
method for calculating the Green's functions from the 
set of local self-energies. We find this method to be much more 
powerful in solving systems with spatial variations or inhomogeneity, and 
it is also faster in bulk systems with current flow than a more 
conventional ${\bf k}$-space integral technique. 

Since the stacked planes have translational symmetry within the plane, 
the systems that we study are inhomogeneous in one direction 
only. We choose that direction to be labeled the z-axis, 
which is also the direction of current flow through the Josephson 
junction. The first stage of our method is to convert the 
problem from a three-dimensional system to a one-dimensional system following 
the algorithm of Potthoff and Nolting~\cite{Pott}. 
We perform a Fourier transform within the planes to determine the planar 
indexed Green's functions,
\begin{equation} 
\underline{\underline{G}}_{\alpha \beta}
(i\omega_{n},k_{x},k_{y}) = \left(\frac{a}{L}\right)^{2}\sum_{x_{j},y_{j}}
\underline{\underline{G}}({\bf r}_{i},{\bf r}_{j},i\omega_{n}) 
\exp \left[ k_{x}(x_{j}-x_{i}) + 
k_{y}(y_{j}-y_{i}) \right],
\end{equation} 
where $\alpha$ and $\beta$ denote 
distinct planes, defined by $\alpha = z_{i}/a$, $\beta=z_{j}/a$
 and the summation is over all 
lattice sites, $(x_{j},y_{j})$, within the $\beta$-th plane. The self energy, 
$\underline{\underline{\Sigma}}_{\alpha}(i\omega_{n}) 
=\underline{\underline{\Sigma}}(z_{i},i\omega_{n})
= \underline{\underline{\Sigma}}({\bf r}_{i},i\omega_{n})$, 
is independent of the planar 
coordinates $x_{i}$ and $y_{i}$, so that Dyson's equation 
[Eq.~(\ref{eq:Dys})] becomes  
\begin{equation}
\underline{\underline{G}}_{\alpha \beta}(i\omega_{n},k_{x},k_{y}) = 
\underline{\underline{G}}^{(0)}_{\alpha \beta}(i\omega_{n},k_{x},k_{y})
+ \sum_{\gamma} \underline{\underline{G}}^{(0)}_{\alpha \gamma}
(i\omega_{n},k_{x},k_{y}) 
\underline{\underline{\Sigma}}_{\gamma} (i\omega_{n})
\underline{\underline{G}}_{\gamma \beta}(i\omega_{n},k_{x},k_{y}),
\end{equation}
with the summation over all planes, $\gamma$. The non-interacting 
planar Green's function, is similarly found by the Fourier transform 
\begin{equation}
\underline{\underline{G}}^{(0)}_{\alpha \beta}(i\omega_{n},k_{x},k_{y}) = 
\left(\frac{a}{L}\right)^{2}\sum_{x_{j},y_{j}}
\underline{\underline{G}}^{(0)}({\bf r}_{i},{\bf r}_{j},i\omega_{n}) 
\exp \left[ k_{x}(x_{j}-x_{i}) + 
k_{y}(y_{j}-y_{i}) \right].
\end{equation} 

The local Green's function, 
$\underline{\underline{G}}({\bf r}_{i},{\bf r}_{i},i\omega_{n})$, 
is required for calculating the self-consistent potentials, 
and the Green's 
function $\underline{\underline{G}}({\bf r}_{i},{\bf r}_{j},i\omega_{n})$ 
between two neighboring sites, ${\bf r}_{i}=(x_{i},y_{i},z_{i})$ and 
${\bf r}_{j}=(x_{i},y_{i},z_{i} \pm a)$ 
is required for current calculations. These are given by the simple 
planar momentum integrals: 
\begin{equation}
\underline{\underline{G}}({\bf r}_{i},{\bf r}_{j},i\omega_{n}) 
= \left(\frac{\pi}{a}\right)^{2} \int^{\pi /a}_{-\pi /a} 
\int^{\pi /a}_{-\pi /a}
\underline{\underline{G}}_{\alpha,\beta}(i\omega_{n},k_{x},k_{y})
dk_{x} dk_{y}
\end{equation}
where again, $\alpha=z_{i}/a$ and $\beta=z_{j}/a$, and the phase factors in the 
integral have canceled as $x_{i}=x_{j}$ and $y_{i}=y_{j}$. 
Hence our goal is to 
find the interacting Green's functions, 
$\underline{\underline{G}}_{\alpha \beta}(i\omega_{n},k_{x},k_{y})$. 

We make a huge improvement (by one to two orders of magnitude) 
in the computational 
efficiency by transforming the two-dimensional planar momentum integral 
into a single integral over in-plane kinetic energy. In the case of 
nearest-neighbor hopping on a square lattice, the 
kinetic energy within the $\alpha$-th plane is given by 
\begin{equation}
\varepsilon^{xy}_{\alpha} = -2t_{\alpha} 
\left[ \cos(k_{x}a) + \cos (k_{y}a)\right]
= \frac{t_{\alpha}}{t} \overline{\varepsilon^{xy}}
\end{equation} 
where $t_{\alpha}$ is the hopping integral between two  nearest-neighbor sites 
within the $\alpha$-th plane and $a$ is the lattice 
spacing. The 
effect of the in-plane kinetic energy is equivalent to an increase in the 
on-site energy, 
$\epsilon^{(0)}_{i} \mapsto \epsilon^{(0)}_{i} + \varepsilon^{(xy)}_{\alpha}$ 
which can vary between the different planes. 
The planar Green's functions only depend on the planar momentum 
via the normalized kinetic energy, 
$\overline{\varepsilon^{xy}} = -2t\left[ \cos(k_{x}a) + \cos (k_{y}a)\right]$, 
such that $\underline{\underline{G}}_{\alpha\beta}(i\omega_{n},k_{x},k_{y})
= \underline{\underline{G}}_{\alpha\beta}
(i\omega_{n},\overline{\varepsilon^{xy}})$. 
Hence, by using the two-dimensional density of 
states, $ \rho^{2D}(\varepsilon)$, for a square lattice, the 
momentum integral is transformed into 
\begin{equation}
 \left(\frac{a}{2\pi} \right)^{2} \int^{\pi/a}_{-\pi/a} \int^{\pi/a}_{-\pi/a}
 G_{\alpha \beta}(i\omega_{n},k_{x},k_{y})dk_{x} dk_{y} 
= \int^{\infty}_{-\infty} G_{\alpha\beta}
(i\omega_{n},\overline{\varepsilon^{xy}})
\rho^{2D}(\overline{\varepsilon^{xy}}) d\overline{\varepsilon^{xy}} . 
\end{equation}
The specific hopping integral, $t_{\alpha}$, for each plane affects the 
on-site potential of a given plane in the continued fraction method 
[see Eq.~(\ref{eq:amatrix}) below], 
but does not contribute to the change of variables in the 
momentum integral. 

Once 
the system is converted to a one-dimensional model, with nearest-neighbor 
hopping, the 
Green's functions can be solved rapidly by a continued fraction expansion, 
without recourse to another $k$-space integral for the $z$-direction. 
The continued fraction expansion 
is similar to the recursion method\cite{Hayd,Hayd2}, modified to include 
superconductivity~\cite{Litak}, but 
with three important differences. First, the method is much faster, as 
there is no need to expand about a site to obtain a new basis --- the system is 
already one-dimensional in form. 
Second, there is no inaccuracy in the termination 
process, as the hopping integrals are given exactly in the model. The third 
point is an alteration, because the sites of interest are not at the end of a 
chain, but in the middle. This leads to a different set of continued fractions 
that must be calculated compared with the standard recursion method. In test 
runs, our method proves to be $4\times 10^{5}$ times faster, and with 
machine precision accuracy, compared to a standard recursion method 
expansion which is terminated (due to memory limits) at 
an accuracy of one part in $10^{3}$! 
An alternate 
approach is to solve Dyson's matrix equation directly for a finite 
system, where the infinitely extended 
bulk boundaries can be mimicked by appropriate choice 
of potentials for the end planes. 
We have carried out such an approach as a 
comparison, but find it to be much slower 
(by a factor of 4000 for 60 planes than our method, and it 
grows like $N^{3}$ for large systems with $N$ planes) so 
we only describe the 
continued-fraction method below. 

The equivalence of our method to the recursion method is that we calculate 
the Green's functions directly from a continued-fraction representation of 
the inverse of the Hamiltonian matrix in real space. 
That is, we find
\begin{equation}
\underline{\underline{G}}_{\alpha \beta}
(i\omega_{n},\overline{\varepsilon^{xy}}) = 
\left( \begin{array}{ccccccc}
\ddots & \ddots  & 0 &\vdots  & \vdots & \vdots &  \\*
\ddots & i\omega_{n} \underline{\underline{1}} -
\underline{\underline{a}}_{\alpha-2} & 
\underline{\underline{b}}_{\alpha-1} & 0 & 0 &
0 & \ldots  \\*
\ldots & \underline{\underline{b}}^{\dagger}_{\alpha-1} &
i\omega_{n} \underline{\underline{1}} -
\underline{\underline{a}}_{\alpha-1} & 
\underline{\underline{b}}_{\alpha}  & 0 & 0 &
\ldots  \\*
\ldots & 0 &  \underline{\underline{b}}^{\dagger}_{\alpha} &
i\omega_{n} \underline{\underline{1}} -
 \underline{\underline{a}}_{\alpha} & 
\underline{\underline{b}}_{n+1}  & 0 &
\ldots  \\*
\ldots & 0 & 0 &  \underline{\underline{b}}^{\dagger}_{\alpha+1} &
i\omega_{n} \underline{\underline{1}} -
 \underline{\underline{a}}_{\alpha+1} & 
\underline{\underline{b}}_{\alpha+2}  & 0 \\*
 & \vdots & \vdots & 0 & \ddots & \mbox{  } \ddots \mbox{  } & 
\mbox{  } \ddots \mbox{  }
\end{array} \right)^{\mbox{\Large{-1}}}_{\mbox{\Large{$\alpha \beta$}}}
\end{equation}
where the matrices $\left\{ \underline{\underline{a}}_{\alpha} \right\}$ are 
the total in-plane energies for a particular plane, given by 
\begin{equation}
\underline{\underline{a}}_{\alpha} = 
\left( \begin{array}{cc}
\epsilon_{\alpha} + \varepsilon^{(xy)}_{\alpha} + 
\Sigma_{\alpha}(i\omega_{n}) - \mu 
& \phi_{\alpha}(i\omega_{n}) \\*
\phi_{\alpha}^{*}(i\omega_{n}) & 
-\epsilon_{\alpha} - \varepsilon^{(xy)}_{\alpha} - 
\Sigma_{\alpha}^{*}(i\omega_{n}) + \mu 
\end{array} \right)
\label{eq:amatrix}
\end{equation}
and $\left\{ \underline{\underline{b}}_{\alpha} \right\}$ 
couple the $\alpha-1$th and 
$\alpha$th planes,
\begin{equation}
\underline{\underline{b}}_{\alpha} = 
\left( \begin{array}{cc}
-t_{\alpha-1,\alpha} & 0 \\* 0 & 
t^{*}_{\alpha-1,\alpha}
\end{array} \right) .
\end{equation}
The local planar Green's function, 
$\underline{\underline{G}}_{\alpha\alpha}(i\omega_{n},
\overline{\varepsilon^{xy}})$, is 
readily evaluated as a combination of continued fractions
(as in the renormalized perturbation expansion~\cite{Econ}). We define the 
right-directed, $\underline{\underline{R}}_{\alpha}(i\omega_{n})$, 
and left-directed $\underline{\underline{L}}_{\alpha}(i\omega_{n})$ 
continued fractions from a plane, $\alpha$, 
recursively as
\begin{eqnarray}
\underline{\underline{R}}_{\alpha}(i\omega_{n},\overline{\varepsilon^{xy}}) 
& = & i\omega_{n}
\underline{\underline{1}} - \underline{\underline{a}}_{\alpha} 
-\underline{\underline{b}}_{\alpha+1} 
\underline{\underline{R}}^{-1}_{\alpha+1}(i\omega_{n},
\overline{\varepsilon^{xy}})
\underline{\underline{b}}^{\dagger}_{\alpha+1} 
\label{eq:right}\\* 
\underline{\underline{L}}_{\alpha}(i\omega_{n},\overline{\varepsilon^{xy}}) 
& = & i\omega_{n}
\underline{\underline{1}} - \underline{\underline{a}}_{\alpha} 
-\underline{\underline{b}}^{\dagger} _{\alpha}
\underline{\underline{L}}^{-1}_{\alpha-1}(i\omega_{n},
\overline{\varepsilon^{xy}})
\underline{\underline{b}}_{\alpha} 
\label{eq:left}
\end{eqnarray}
The recursive calculation continues to infinity, but once it 
has been extended to planes in the uniform bulk medium, 
where $\alpha<1-N/2$ or $\alpha>N/2$, the coefficients at 
each level become constant. The effect of a constant phase gradient in 
$\phi$ is equivalent to a constant phase factor in the hopping integral, 
$t_{\alpha,\alpha+1}$, 
that does not change between planes in the bulk. Hence, by equating all 
$\underline{\underline{R}}_{\alpha}(i\omega_{n},
\overline{\varepsilon^{xy}})$ as 
$\underline{\underline{R}}_{\infty}(i\omega_{n},
\overline{\varepsilon^{xy}})$ for $\alpha>N/2$ and 
$\underline{\underline{L}}_{\alpha}(i\omega_{n},
\overline{\varepsilon^{xy}})$ as 
$\underline{\underline{L}}_{\infty}(i\omega_{n},
\overline{\varepsilon^{xy}})$ for $\alpha<1-N/2$ 
in the bulk limit, an exact 
terminator function can be calculated as the solution of a complex 
quadratic matrix equation:
\begin{eqnarray}
\underline{\underline{R}}_{\infty}(i\omega_{n},\overline{\varepsilon^{xy}})
\underline{\underline{b}}_{\infty}^{\dagger -1}
\underline{\underline{R}}_{\infty}(i\omega_{n},\overline{\varepsilon^{xy}}) 
+ \left[ \underline{\underline{a}}_{\infty} - 
i\omega_{n}\underline{\underline{1}}\right] 
\underline{\underline{b}}_{\infty}^{\dagger -1}
\underline{\underline{R}}_{\infty}(i\omega_{n},\overline{\varepsilon^{xy}}) + 
\underline{\underline{b}}_{\infty} & = & \underline{\underline{0}}\\
\underline{\underline{L}}_{\infty}(i\omega_{n},\overline{\varepsilon^{xy}})
\underline{\underline{b}}_{\infty}^{-1}
\underline{\underline{L}}_{\infty}(i\omega_{n},\overline{\varepsilon^{xy}}) 
+ \left[ \underline{\underline{a}}_{\infty} - 
i\omega_{n}\underline{\underline{1}}\right] 
\underline{\underline{b}}_{\infty}^{-1}
\underline{\underline{L}}_{\infty}(i\omega_{n},\overline{\varepsilon^{xy}}) + 
\underline{\underline{b}}^{\dagger}_{\infty} & = & \underline{\underline{0}} .
\end{eqnarray}
Note that the same terminator 
function is used for all sites in the intermediate layers, and the functions 
$\underline{\underline{R}}_{\alpha}$ and 
$\underline{\underline{L}}_{\alpha}$ calculated for 
one site are used in the calculation for the next site --- so the number of 
computations required to find solutions for all sites can be less than $O(N)$. 

There are two ways to solve this matrix quadratic equation. When we perform 
calculations on the real axis, without including a supercurrent, the 
matrix equation becomes analytically tractable to solve. On the imaginary 
axis, we find that it is numerically faster to simply find an iterative solution 
to these quadratic equations within the bulk medium. In most cases, accuracies 
of one part in $10^{10}$ can be achieved in ten iterations or less.

The continued fractions form the local planar Green's 
functions, according to
\begin{equation}
\underline{\underline{G}}_{\alpha\alpha}(i\omega_{n},
\overline{\varepsilon^{xy}})
= \left\{ i\omega_{n}\underline{\underline{1}}
- \underline{\underline{a}}_{\alpha}(\overline{\varepsilon^{xy}})
- \underline{\underline{b}}^{\dagger} _{\alpha}
\underline{\underline{L}}^{-1}_{\alpha-1}(i\omega_{n},
\overline{\varepsilon^{xy}})
\underline{\underline{b}}_{\alpha} 
-\underline{\underline{b}}_{\alpha+1} 
\underline{\underline{R}}^{-1}_{\alpha+1}(i\omega_{n},\overline{\varepsilon^{xy}})
\underline{\underline{b}}^{\dagger}_{\alpha+1} \right\}^{-1}
\end{equation}
which, using Eqs.(\ref{eq:right}-\ref{eq:left}), can be simplified to 
\begin{equation}
\underline{\underline{G}}_{\alpha\alpha}(i\omega_{n},
\overline{\varepsilon^{xy}})
= \left[\underline{\underline{R}}_{\alpha} + \underline{\underline{L}}_{\alpha}
- i\omega_{n}\underline{\underline{1}} + \underline{\underline{a}}_{\alpha} 
\right]^{-1}.
\end{equation}
The Green's functions connecting neighboring planes, $\alpha$ and 
$\alpha \pm 1$, which 
are required to calculate the current flow, are 
given in two equivalent forms 
\begin{eqnarray}
\underline{\underline{G}}_{\alpha,\alpha+1}(i\omega_{n},
\overline{\varepsilon^{xy}}) &
= & -\underline{\underline{G}}_{\alpha\alpha}(i\omega_{n},
\overline{\varepsilon^{xy}})
\underline{\underline{b}}_{\alpha+1} 
\underline{\underline{R}}^{-1}_{\alpha+1}(i\omega_{n},
\overline{\varepsilon^{xy}}) 
 =  -\underline{\underline{L}}^{-1}_{\alpha}(i\omega_{n},
\overline{\varepsilon^{xy}})
\underline{\underline{b}}_{\alpha+1} 
\underline{\underline{G}}_{\alpha+1,\alpha+1}(i\omega_{n},
\overline{\varepsilon^{xy}}) \\*
\underline{\underline{G}}_{\alpha,\alpha-1}(i\omega_{n},
\overline{\varepsilon^{xy}}) &
= & -\underline{\underline{G}}_{\alpha\alpha}(i\omega_{n},
\overline{\varepsilon^{xy}})
\underline{\underline{b}}^{\dagger} _{\alpha}
\underline{\underline{L}}^{-1}_{\alpha-1}(i\omega_{n},
\overline{\varepsilon^{xy}}) 
 =  -\underline{\underline{R}}^{-1}_{\alpha+1}(i\omega_{n},
\overline{\varepsilon^{xy}})
\underline{\underline{b}}^{\dagger} _{\alpha}
\underline{\underline{G}}_{\alpha-1,\alpha-1}
(i\omega_{n},\overline{\varepsilon^{xy}}) 
\end{eqnarray}
The local planar Green's functions enable us to calculate the self-energy and 
electron density, $n_{i}$, on each site in a given plane, $\alpha=z_{i}/a$, 
with the latter given by
\begin{equation}
n_{i} = k_{B}T\sum_{\omega_{n}} 
\int^{\infty}_{-\infty}\rho^{2D}(\overline{\varepsilon^{xy}}) 
\mbox{\large Im} \left[ G_{\alpha\alpha}(i\omega_{n},
\overline{\varepsilon^{xy}}) 
\right]d\overline{\varepsilon^{xy}} .
\end{equation}
The current, $I_{\alpha,\alpha+1}$, 
which flows along each link between two neighboring 
planes, $\alpha$ and $\alpha+1$, in the $z$-direction is given by:
\begin{equation}
I_{\alpha,\alpha+1} = \frac{2te}{\hbar}k_{B}T\sum_{\omega_{n}} 
\int^{\infty}_{-\infty}\rho^{2D}(\overline{\varepsilon^{xy}}) 
\mbox{\large Im} \left[ G_{\alpha,\alpha+1}(i\omega_{n},
\overline{\varepsilon^{xy}}) 
\right] d\overline{\varepsilon^{xy}} .
\end{equation}
A stringent convergence check for self-consistency, when there is 
a phase difference between the bulk superconductors, 
is that the current flow is constant from one plane to the next. 

For the bulk boundary regions, 
the uniform variation of phase in the off-diagonal self energy, 
$\phi({\bf r}_{i})$ has the form, 
$\phi({\bf r}_{i}) = \phi_{0} \exp\left[i{\bf q}\cdot{\bf r}_{i}\right]$, 
where the net superfluid momentum depends on ${\bf q}=(0,0,q_{z})$ through 
$mv_{s} = \frac{\hbar}{a}\sin(q_{z}a)$ and 
leads to the following solution of Dyson's equation:
\begin{equation} 
G({\bf r}_{i},{\bf r}_{j},i\omega_{n}) = \int d^{3}{\bf k} 
\frac{ \left( i\omega_{n} + \varepsilon_{{\bf k}-{\bf q}} - \mu + 
\Sigma^{*}(i\omega_{n}) 
\right) \mbox{\rm e}^{i {\bf k}\cdot({\bf r}_{i}-{\bf r}_{j}) } }
{ \left( i\omega_{n} -\varepsilon_{{\bf k}} + \mu - \Sigma(i\omega_{n}) \right)
\left( i\omega_{n} + \varepsilon_{{\bf k}-{\bf q}} - \mu + 
\Sigma^{*}(i\omega_{n})\right)
- \left| \phi_{0}(i\omega_{n}) \right|^{2} }
\label{eq:greenij}
\end{equation}
and 
\begin{equation} 
F({\bf r}_{i},{\bf r}_{j},i\omega_{n}) = \int d^{3}{\bf k} 
\frac{ \phi_{0}(i\omega_{n})
 \mbox{\rm e}^{i ({\bf k}-{\bf q})\cdot({\bf r}_{i}-{\bf r}_{j}) } }
{ \left( i\omega_{n} -\varepsilon_{{\bf k}} + \mu - \Sigma(i\omega_{n}) \right)
\left( i\omega_{n} + \varepsilon_{{\bf k}-{\bf q}} - \mu + 
\Sigma^{*}(i\omega_{n})\right)
- \left| \phi_{0}(i\omega_{n}) \right|^{2} }
\end{equation}
where the diagonal self energy, $\Sigma(i\omega_{n})$, is independent of 
site index, ${\bf r}_{i}$, in the bulk. We will need only the local, and 
nearest-neighbor bulk Green's functions in our calculations.

\subsection{Local self energy calculations}

In this contribution, the superconducting region is 
modeled by the negative-U Hubbard model within the Hartree-Fock 
(static mean-field) 
approximation. In this case, the local self energy is found from 
the local Green's functions by:
\begin{equation}
\Sigma({\bf r}_{i},i\omega_{n}) = U T \sum_{\omega_{n}} 
G({\bf r}_{i},{\bf r}_{i},i\omega_{n})
\label{eq:selfa}
\end{equation}
 and
\begin{equation}
\phi({\bf r}_{i},i\omega_{n}) = -U T \sum_{\omega_{n}} 
F({\bf r}_{i},{\bf r}_{i},i\omega_{n}),
\label{eq:selfb}
\end{equation}
where the instantaneous 
electron-electron interaction energy, $U$, leads to a time-independent self 
energy. This procedure is identical to the conventional 
Bogoliubov-de Gennes approach~\cite{DeGe}, 
which neglects retardation effects in the 
superconductor. As all sites within a plane are identical, the 
self energy need only be calculated once for each of the $N$ planes.

\begin{figure}
  \centerline{\psfig{figure=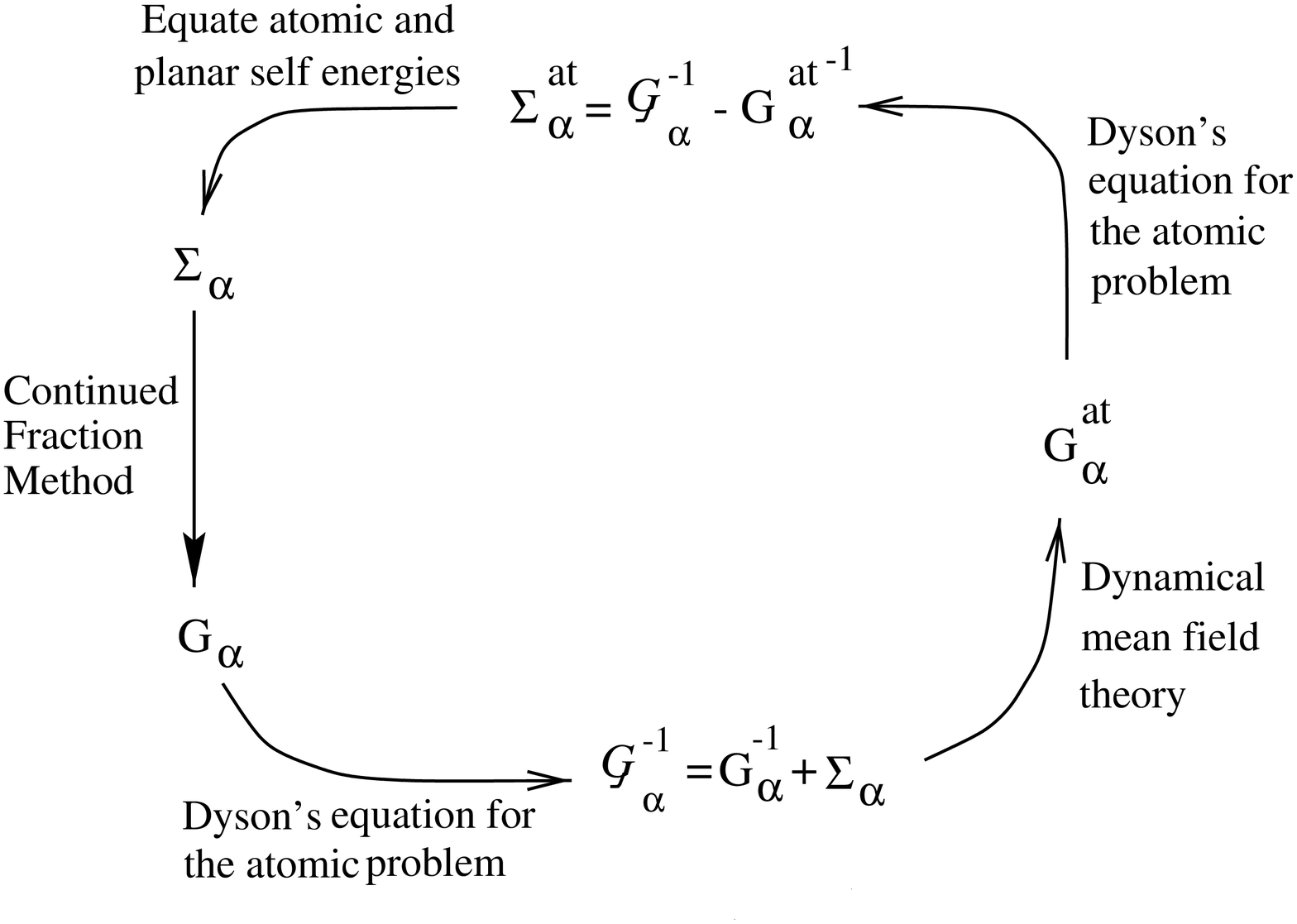,width=7.0in}}
  \caption{Diagram of the iteration procedure, where the dynamical mean 
field theory is used to calculate local self-energies from local 
Green's functions. In the case of the Hartree-Fock approximation for the self 
energy, the dynamical mean field theory step is trivial, 
[see Eqs.~(\ref{eq:selfa}-\ref{eq:selfb})].}
\label{fig:iter}
\end{figure}

For planes within the barrier which include impurities, the dynamical mean 
field approximation says that the local (site) Green's function, 
$\underline{\underline{G}}({\bf r}_{i},{\bf r}_{i},i\omega_{n})$ 
is related to a local host Green's function~\cite{Brandt}, 
$\underline{\underline{\cal{G}}}({\bf r}_{i},i\omega_{n})$, 
via
\begin{equation}
\underline{\underline{\cal{G}}}({\bf r}_{i},i\omega_{n}) = \left[ 
\underline{\underline{G}}^{-1}({\bf r}_{i},{\bf r}_{i},i\omega_{n}) + 
\underline{\underline{\Sigma}}({\bf r}_{i},i\omega_{n})\right]^{-1}.
\label{eq:itera}
\end{equation}
The atomic Green's function, 
$\underline{\underline{G^{at}}}({\bf r}_{i},i\omega_{n})$, 
which will be equated to the local Green's function, 
$\underline{\underline{G}}({\bf r}_{i},{\bf r}_{i},i\omega_{n})$, 
in the dynamical mean field approximation, then satisfies 
\begin{equation}
\underline{\underline{G^{at}}}({\bf r}_{i},i\omega_{n}) = 
(1-\rho_{imp})\underline{\underline{\cal{G}}}({\bf r}_{i},i\omega_{n}) + 
\rho_{imp} 
\left[ \underline{\underline{\cal{G}}}^{-1}({\bf r}_{i},i\omega_{n})
- U^{FK} \underline{\underline{1}} \right]^{-1}.
\label{eq:iterb}
\end{equation}
and the local self energy becomes 
\begin{equation}
\underline{\underline{\Sigma}}({\bf r}_{i},i\omega_{n}) = 
\underline{\underline{\cal{G}}}^{-1}({\bf r}_{i},i\omega_{n})
- \underline{\underline{G^{at}}}^{-1}({\bf r}_{i},i\omega_{n}).
\label{eq:iterc}
\end{equation}
Starting from a local self energy, 
$\underline{\underline{\Sigma}}({\bf r}_{i},i\omega_{n})$, and a local 
Green's function, 
$\underline{\underline{G}}({\bf r}_{i},{\bf r}_{i},i\omega_{n})$, 
Eqs.(\ref{eq:itera},\ref{eq:iterb},\ref{eq:iterc}) can be employed 
to iteratively determine a new self energy 
$\underline{\underline{\Sigma}}({\bf r}_{i},i\omega_{n})$ 
when the plane is described by the Falicov-Kimball model. The method, 
which is solved for a fixed concentration of impurities, $\rho_{imp}$, is 
equivalent to the coherent potential approximation. The algorithm is 
summarized in Fig.~\ref{fig:iter}.

\section{Phase Variation}

Standard theory of Josephson junctions~\cite{Jose} 
predicts the phase variation of the 
current in the weak-coupling limit to be $I(\theta) = I_{c} \sin (\theta)$ 
where $\theta$ is the phase difference across the barrier, and $I_{c}$ is 
the temperature-dependent critical current. Such a current variation arises 
from consideration of two superconductors with different phases being 
coupled by a single energy-independent 
transmission coefficient, corresponding to the tunneling 
of Cooper pairs across a barrier. A more general 
consideration~\cite{Furu} includes the 
Bogoliubov-de Gennes equations for a one-dimensional system of two 
superconductors 
coupled across a potential barrier, $V$. The strength of the barrier is 
measured by $Z=mV/\hbar^{2}k_{F}$ which determines the transmission 
coefficient, $\tau$, according to ${\Large \tau}=1/(1+Z^{2})$. It is found that 
Andreev bound 
states carry the current, whose variation with phase difference across the 
barrier leads to~\cite{Furu} 
\begin{equation}
I = -G_{N} \frac{\pi \Delta_{0} }{2e} \frac{ \sin (\theta) }
{\sqrt{1 - {\Large \tau} \sin^{2}(\theta/2) } } 
\tanh \left( \frac{ \Delta_{0} \cos (\theta /2) }{2k_{B} T}\right).
\end{equation}
This reproduces the weak-coupling result~\cite{Ambe} in the limit 
of low transparency, ${\Large \tau} \ll 1$, 
with 
\begin{equation}
I = -G_{N} \frac{\pi \Delta_{0} }{2e} \sin (\theta) 
\tanh \left( \frac{ \Delta_{0} }{2k_{B} T} \right),
\end{equation}
and reproduces the formula for a point contact Josephson junction 
with perfect transparency~\cite{Kuli}, ${\Large \tau} = 1$:
\begin{equation}
I = -G_{N} \frac{\pi \Delta_{0} }{e}  \sin (\theta/2)
\tanh \left( \frac{ \Delta_{0} \cos (\theta /2) }{2k_{B} T} \right).
\end{equation}
Interestingly, the maximum current is at a phase difference which approaches 
$\theta = \pi$ at low temperatures in the latter case, whereas traditional 
tunnel junctions have their maximum at $\theta =\pi /2$.

In the 
more general case of a three-dimensional geometry~\cite{Tana}, the 
transmissivity depends on the angle of approach of the electron. 
The full current is obtained 
by an angular integral, $d\alpha$ over the Fermi surface. 
The complete calculation leads to:
\begin{equation}
I\left( \theta \right) = -G_{N}\frac{\pi\Delta_{0}(T)}{2e}
\cdot\overline{R}_{N} \sin(\theta)
\int^{\pi/2}_{-\pi/2} \frac{ \sigma_{N}(\alpha) \cos(\alpha) }
{\sqrt{ 1 - \sigma_{N}(\alpha) \sin^{2}(\theta /2) } }
\tanh \left( \frac{ \Delta_{0}(T) 
\sqrt{ 1 - \sigma_{N}(\alpha) \sin^{2}(\theta /2) } }{2k_{B} T} \right)
d\alpha
\end{equation}
where 
\begin{equation}
\overline{R}_{N}^{-1} = \int^{\pi/2}_{-\pi/2}
\sigma_{N}(\alpha) \cos(\alpha) d\alpha .
\end{equation}
The transmission probability for a single electron, which is 
proportional to its contribution to the conductivity, $\sigma_{N}$, 
(from a Landauer formula) 
depends on its angle of incidence, as well as the strength of scattering at 
the barrier. 

The linear response of current due to a small phase difference, $I'$, 
is given by:
\begin{equation}
I' = \left( \frac{\partial I}{\partial \theta} \right)_{\lim \theta \mapsto 0}
= -G_{N} \frac{\pi \Delta_{0}(T) }{2e} 
\tanh \left( \frac{ \Delta_{0}(T) }{2k_{B} T} \right)
\label{eq:linearphase}
\end{equation}
for both the one-dimensional and three-dimensional cases above. 
So if the barrier is parameterized by a single scattering 
potential, then the linear response supercurrent only depends on the 
normal state conductance according to Eq.~(\ref{eq:linearphase}) and the 
product of $I'\cdot R_{N}$ is a constant, independent of the microscopic 
details of the barrier. More 
realistic treatments of the barrier region are expected to result in 
deviations from Eq.~(\ref{eq:linearphase}) and allow $I'\cdot R_{N}$ to 
vary with the properties of the barrier. 

Further differences arise between methods which treat the 
barrier as a single scattering potential and our self-consistent 
treatment of the order parameter and off-diagonal Green's functions within 
the barrier. Our results indicate that the effective 
scattering strength of the barrier, $Z$, changes 
with temperature and current flow. In general, 
the proximity effect, which enhances coupling between superconductors, is 
weakened as the current flow approaches the critical current. Hence at large 
current flow, the effective barrier is increased compared to its value in 
the linear response regime at close to zero current flow. The effect of 
self-consistency on the current-phase relationship appears to be more 
marked for weak barriers, where the current flow becomes relatively large. 

We quantify the current response by 
making use of the linear response 
$I' = \left(dI/d\theta\right)_{lim \theta \mapsto 0}$, 
as well as the maximum current 
flow, $I_{c}$, through the barrier. 
In the weak-coupling limit $I_{c}$ is exactly equal to $I'$, and our results 
show that in general the two are within 20\%
of each other, and scale 
almost 
identically with external parameters. As $I'$ requires much less computational 
time to calculate, than $I_{c}$, we report values of $I'$ for many of 
our results. 

\section{Bogoliubov-de Gennes Results}

To begin, we demonstrate how our model 
reproduces standard results, by using the Hartree-Fock approximation to 
calculate the self energy within the barrier region. As such, we are 
effectively self-consistently solving the full Bogoliubov-de Gennes 
equations for the system. In all the results that we 
present in this paper, the superconducting region is modeled with an 
attractive Hubbard interaction of $U=-2$ at half filling. 
The homogeneous bulk superconductor 
then has a critical temperature, $T_{c} =0.11$ and a zero temperature order 
parameter, $\Delta_{0} = 0.198$.

\subsection{Barrier thickness}

\begin{figure}
  \centerline{\psfig{figure=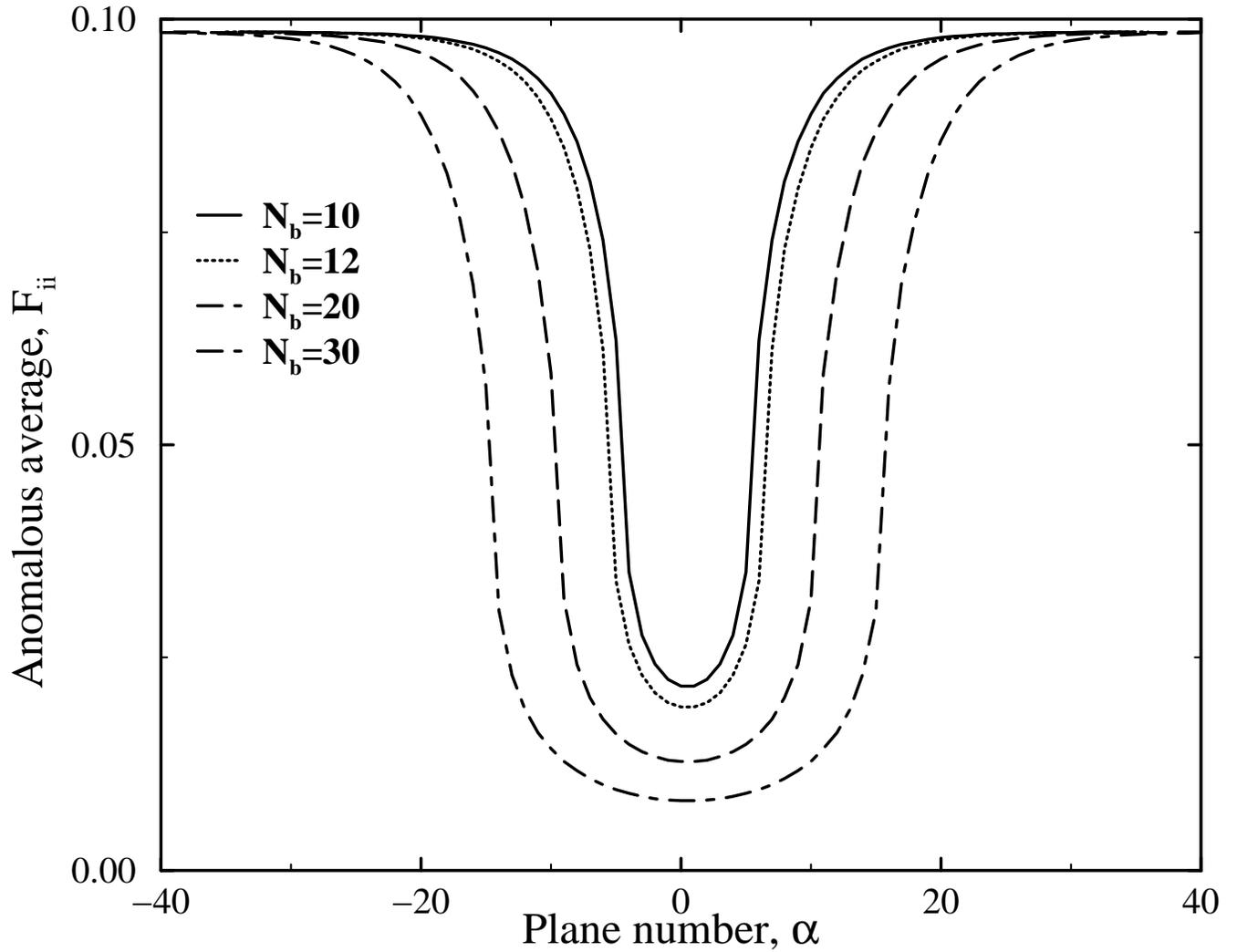,width=7.0in}}
  \caption{ Plot of the decay of the anomalous average 
due to the proximity 
effect in the barrier region. Note that the order parameter, $\Delta$ is 
equal to $2F_{ii}=2F({\bf r}_{i},{\bf r}_{i},\tau=0^{+})$ 
in the superconducting region, where $U=-2$, and is equal to $F_{ii}/2$ in 
the barrier region, as $U_{b}=-1/2$. It is the anomalous Green's function, 
$F_{ii}$, rather than the order parameter, $\Delta_{i}$, that is continuous 
throughout the system. }
\label{fig:widtha}
\end{figure}

\begin{figure}
  \centerline{\psfig{figure=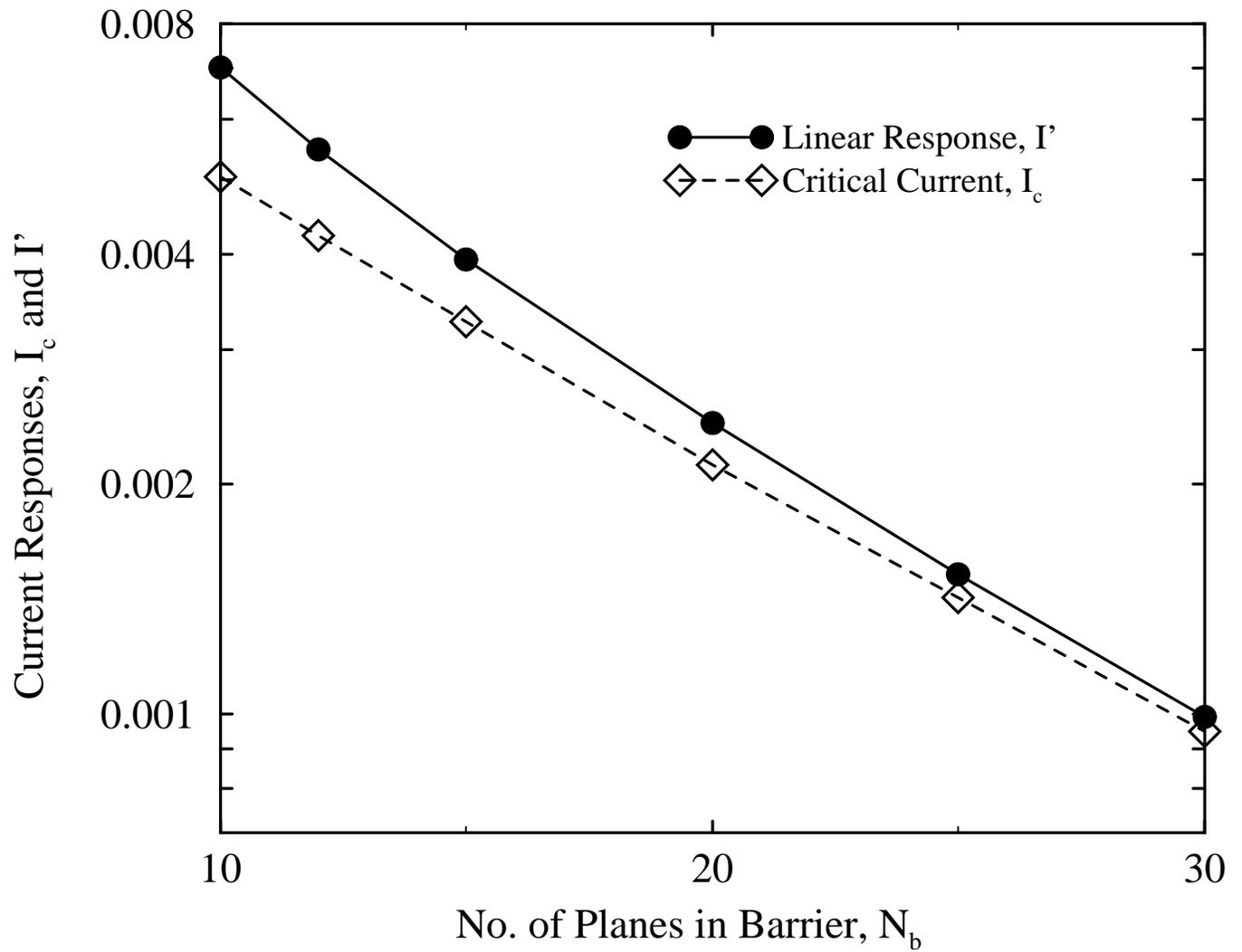,width=7.0in}}
  \caption{Exponential decay of the linear response current, $I'$, and 
critical current, $I_{c}$ with increasing barrier thickness. Note how 
both $I'$ and $I_{c}$ systematically track with each other, and agree 
to within 20\% 
for cases considered here.}
\label{fig:widthb}
\end{figure}

We begin by solving for systems with a small attractive interaction within the 
barrier, $U_{b}=-0.5$, over a range of barrier thicknesses. 
This attraction is small enough that the bulk superconducting transition 
temperature of the barrier material is always less than any temperature we 
consider. 
Fig.~\ref{fig:widtha} demonstrates the proximity effect, 
with the decay of the anomalous average at the 
center of the barrier as its width is increased. 
Fig.~\ref{fig:widthb} is a log plot of the 
linear-response current and critical current against barrier thickness. 
As expected, both the linear-response current, 
$I'$, and the critical current, $I_{c}$, drop rapidly when the number of 
planes 
within the barrier region is increased from 5 to 30. 
The nearly constant slopes indicate that the decays are close to 
exponential. The exponent is approximately twice the bulk correlation length, 
given by 
$2\xi \approx \hbar v_{f}/\Delta \approx 2ta/\Delta \approx 10 a $, where 
$a$ is the lattice spacing.

\begin{figure}
  \centerline{\psfig{figure=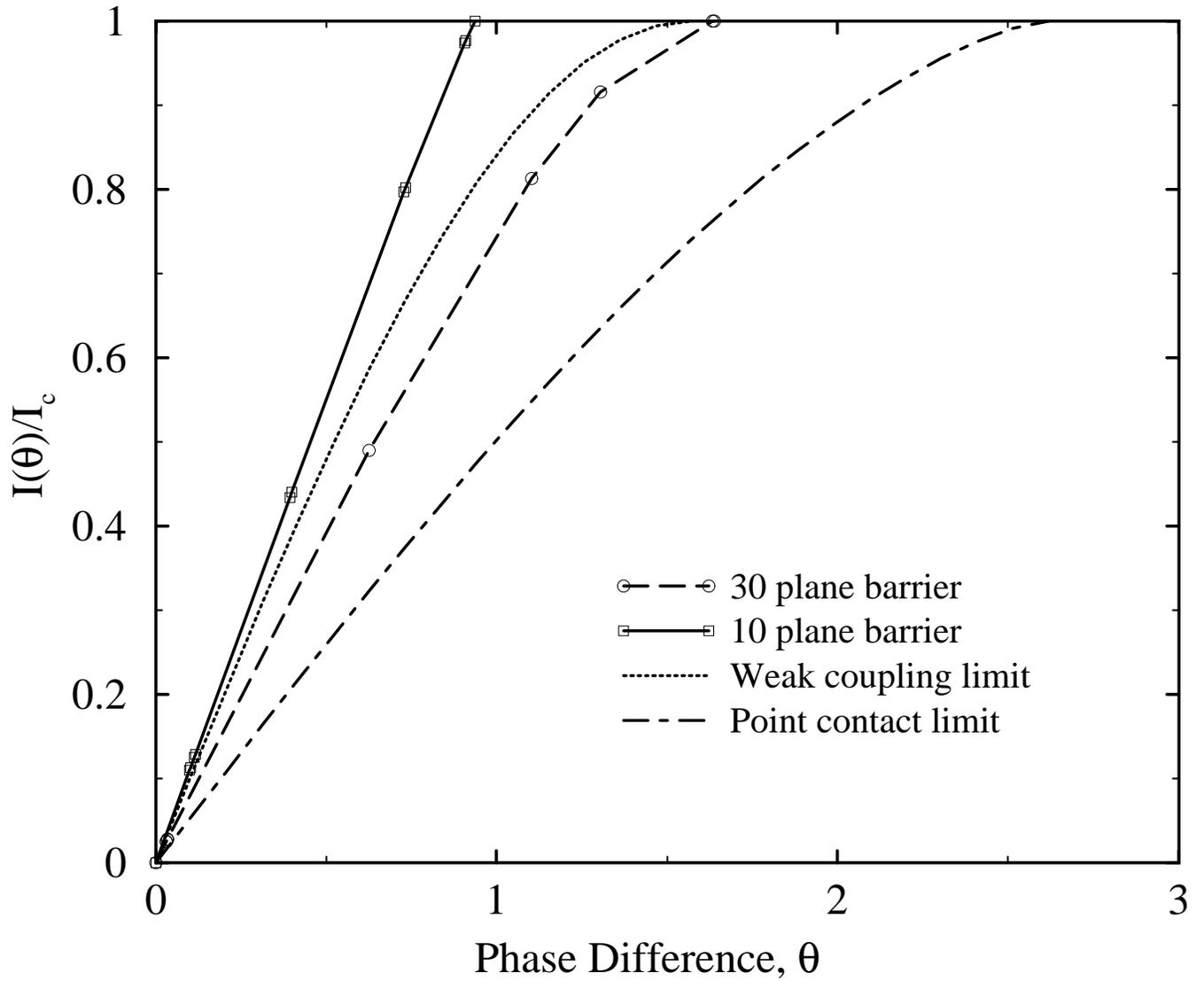,width=7.0in}}
  \caption{Current flow as a function of phase difference across the 
barrier for a thin and a thick junction, compared to analytic results for 
a low transmissivity (dotted line) and a high transmissivity 
(chain-dashed line) junction. Notice how the results for a thin junction 
lie outside of the results for the two analytic limiting cases. This shows 
how including self-consistency and a microscopic model of the barrier 
width can affect the current-phase relation. }
\label{fig:phase}
\end{figure}

Fig.~\ref{fig:phase} shows the normalized 
current as a function of phase difference 
across the barrier region, for two different barrier thicknesses. 
The phase variation is compared to the simple 
form $I(\theta) / I_{c} = \sin(\theta)$, which is appropriate in the 
weak-coupling limit, and 
$I(\theta)/I_{c} = (I_{0}/I_{c})
\sin(\theta/2) \tanh(\Delta\cos(\theta/2) /2k_{B}T)$, 
the result for a point-contact junction, appropriate in the limit of high 
transmissivity. The curve for a thin junction falls outside these two limits, 
indicating that the effects of self-consistency and finite junction width 
are important in determining the current-phase relation for a Josephson 
junction. In particular, the calculated difference between a weakly 
coupled and strongly coupled junction lies in the opposite direction to the 
analytic formula on such a normalized curve. The actual magnitude 
of the critical current, is of course, greatly enhanced for the thinner 
barrier, as shown in Figs.~\ref{fig:widtha}-\ref{fig:widthb}. 
 
\subsection{Barrier interaction strength}

\begin{figure}
  \centerline{\psfig{figure=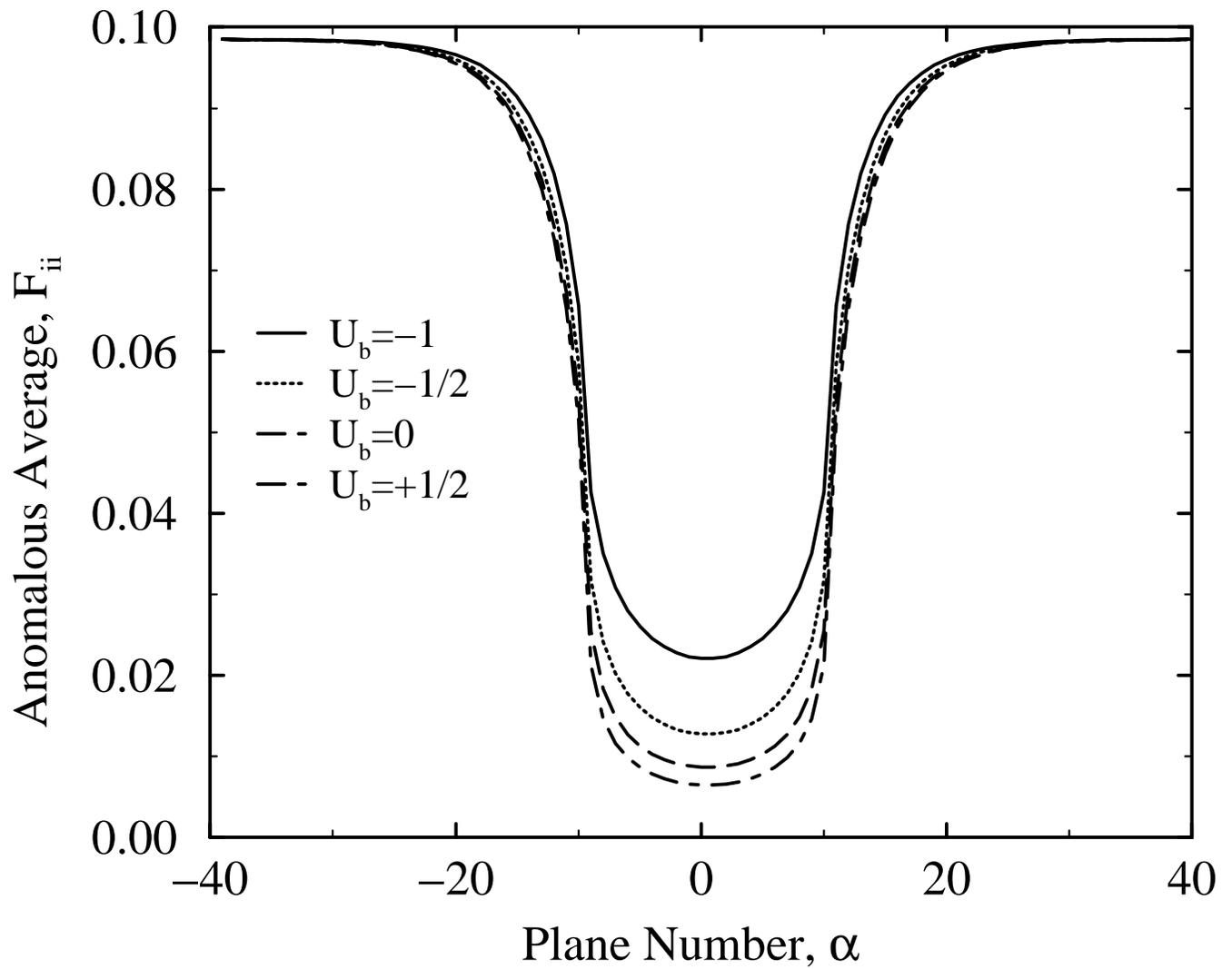,width=7.0in}}
  \caption{ The decay of the anomalous average within the barrier region 
as a function of the Hubbard interaction, $U_{b}$. Note how the anomalous 
average decays more rapidly as the Coulomb interaction in the barrier, 
$U_{b}$, increases in value (and becomes repulsive). }
\label{fig:ubarra}
\end{figure}

\begin{figure}
  \centerline{\psfig{figure=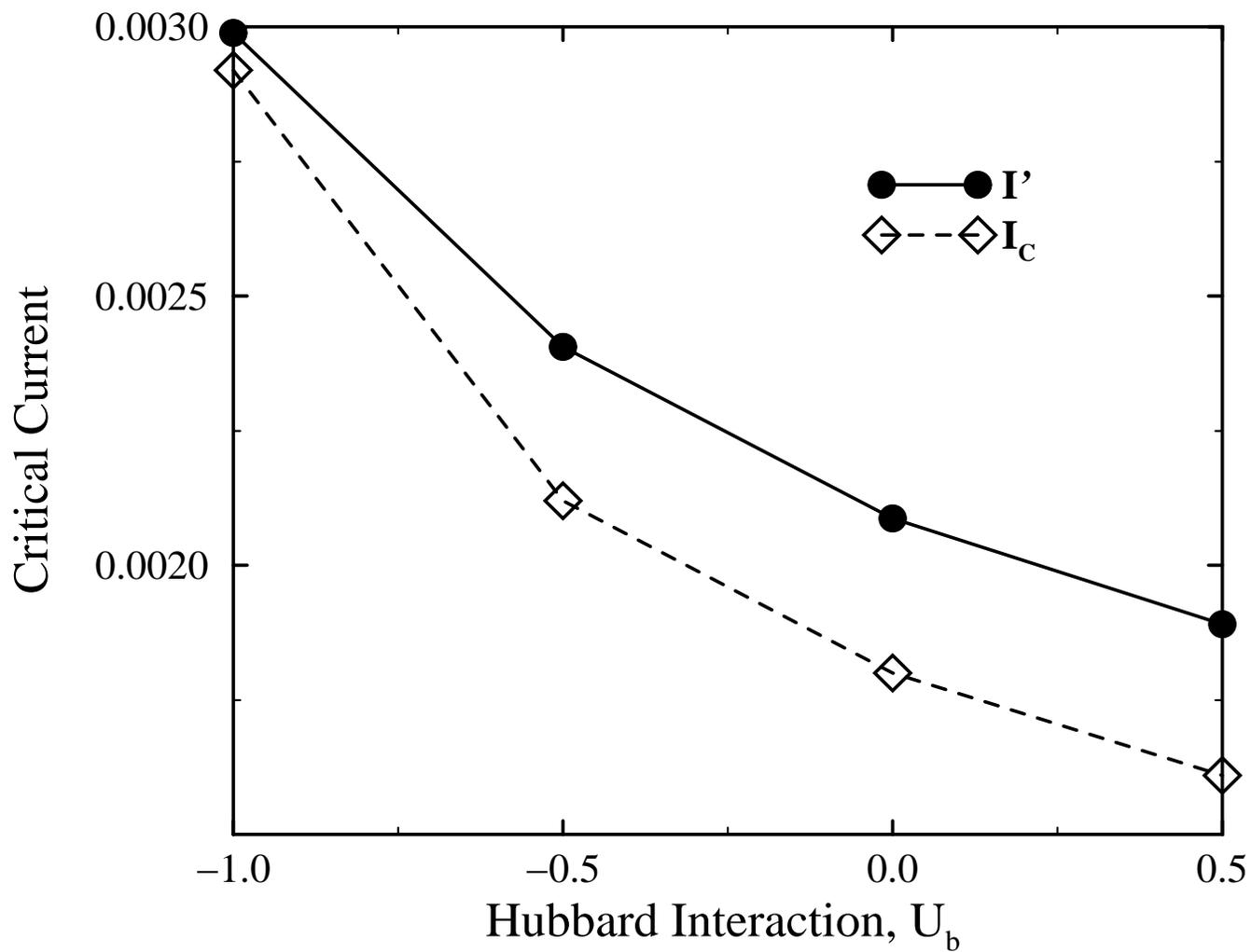,width=7.0in}}
  \caption{ Linear response current, $I'$, and 
critical current, $I_{c}$, as a function of the Coulomb interaction in 
the barrier, $U_{b}$. Note how both critical currents fall with $U_{b}$, 
and note that there 
is no apparent discontinuity through $U_{b}=0$, where the sign of $\Delta$ 
changes within the barrier.  }
\label{fig:ubarrb}
\end{figure}

Fig.~\ref{fig:ubarra} demonstrates the proximity effect as a decay in the 
anomalous average, $F_{ii} = F({\bf r}_{i},{\bf r}_{i},\tau=0^{+})$, 
within a barrier region of twenty planes. 
The proximity effect is seen to depend on the Hubbard interaction, $U_{b}$, 
for sites 
within the barrier. The corresponding critical currents, $I_{c}$ and linear 
response currents, $I'$, are shown in Fig.~\ref{fig:ubarrb}. Note that in the 
cases where $U_{b}$ is negative, the barrier region is actually a 
superconductor in its normal state, above its transition temperature. 
In the case where $U_{b}=0$, the order parameter, $\Delta_{i}$, is exactly 
zero within the barrier. In the example where $U_{b}$ is positive, the order 
parameter actually switches sign within the barrier region, even when there is 
no external phase variation. The results of Fig.~\ref{fig:ubarrb} indicate 
that such a switching in sign of the pairing potential, 
$\Delta_{i}$ has no marked effect on the 
transport properties, which depend on the continuously varying Green's 
functions of the system. Once again we find a systematic tracking of $I'$ and 
$I_{c}$ with the strength of the Coulomb interaction, $U_{b}$, in the barrier. 

\subsection{Barrier hopping integral}

Our method allows us to consider different hopping integrals in different 
regions (within planes or between planes). 
This would be appropriate when the barrier region has a density of 
states at the Fermi surface, or a Fermi velocity that differs from 
that found in the normal 
state of the superconducting regions. When modeling such systems, 
the hopping integral between successive 
planes that have differing intraplanar hopping integrals, is taken as the 
geometric mean of the two planar values 
($t_{\alpha,\alpha+1} = \sqrt{t_{\alpha}.t_{\alpha+1}}$).

\begin{figure}
  \centerline{\psfig{figure=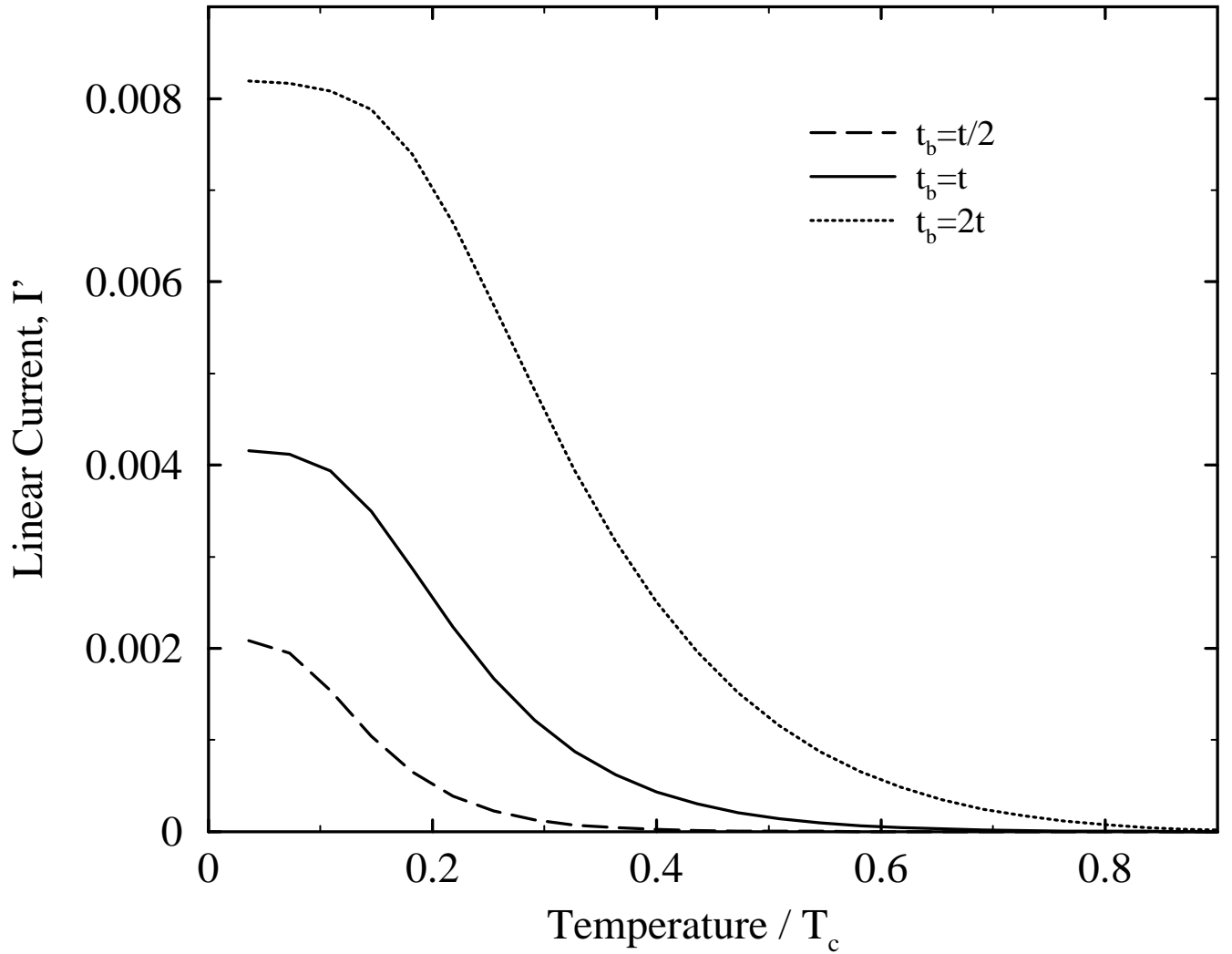,width=7.0in}}
  \caption{ Linear current, $I'$, versus temperature for barriers with 
a hopping integral, $t_{b}$, that can differ from the hopping integral, $t$, 
between planes of the superconductor. Note how $I'$ scales with $t_{b}$ in 
the low-temperature regime. }
\label{fig:thopa}
\end{figure}

\begin{figure}
  \centerline{\psfig{figure=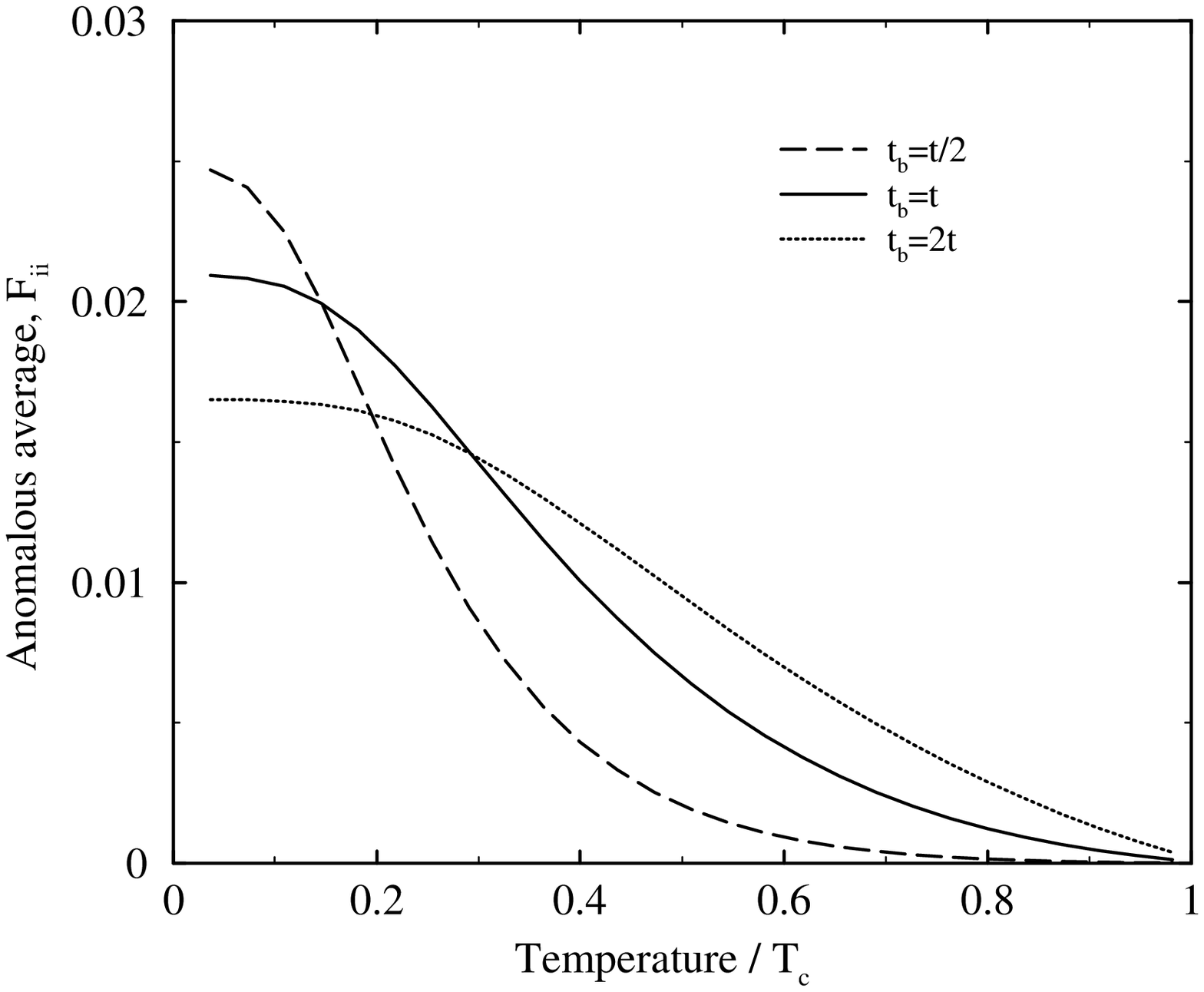,width=7.0in}}
  \caption{ The variation with temperature of the anomalous average, 
$F_{ii}=F({\bf r}_{i},{\bf r}_{i},\tau=0^{+})$, 
for the plane at the center of the barrier region. The different 
curves are results for different values of the barrier hopping integral, 
$t_{b}$. }
\label{fig:thopb}
\end{figure}

\begin{figure}
  \centerline{\psfig{figure=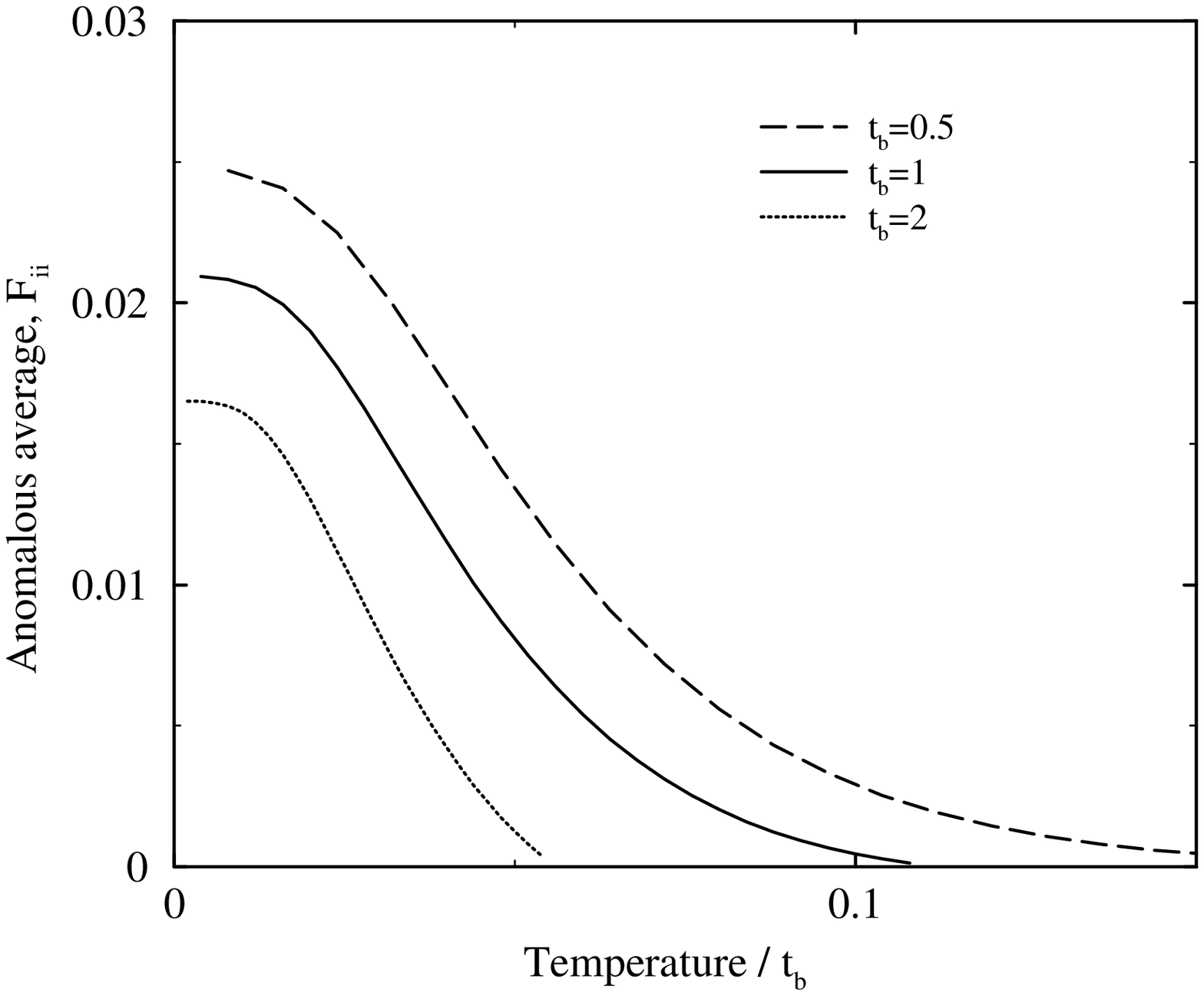,width=7.0in}}
  \caption{ The variation with temperature of the anomalous average, 
$F_{ii}=F({\bf r}_{i},{\bf r}_{i},\tau=0^{+})$, 
for the plane at the center of the barrier region. The different 
curves are results for different values of the barrier hopping integral, 
$t_{b}$, with the temperature normalized by $t_{b}$. }
\label{fig:thopb2}
\end{figure}

Fig.~\ref{fig:thopa} indicates that 
the critical current approximately scales with the hopping integral in the 
barrier region at low temperature. This effect dominates over any increased 
scattering at the interfaces due to Fermi velocity mismatch. 
In fact, when $t_{b}$ is large, the critical current is enhanced by a 
factor of $t_{b}/t$ over that found in the uniform bulk system ($t_{b}=1$). 
The proximity 
effect results in a minimum of the anomalous average, 
$F_{ii}=F({\bf r}_{i},{\bf r}_{i},\tau=0^{+})
=<c_{i\uparrow}(0^{+})c_{i\downarrow}(0)>$, at the 
center of the barrier region. At low temperatures, Fig.~\ref{fig:thopb} 
indicates that the system with a smaller 
hopping integral in the barrier, $t_{b}=1/2$, has a larger anomalous average 
than the uniform system (with $t_{b}=1$), 
due to its increase in density of states at the Fermi surface. As the 
temperature is increased, there is a crossover, with the system that has 
largest hopping integral in the barrier ($t_{b}=2$), having the largest 
anomalous average just below the critical temperature, $T_{c}$. 
Such a crossover is due to the differing natural energy scales, $t_{b}$, 
in the barrier region. For the system with $t_{b}=1/2$, an actual temperature 
of $T=0.1$ corresponds to a temperature of $0.2$ in the natural energy units 
of the barrier, $t_{b}$. 
When the temperatures are given in units of $t_{b}$, as shown in 
Fig.~\ref{fig:thopb2}, 
 a series of approximately parallel curves is seen, ordered according 
to the differing densities of states at the Fermi surface. The anomalous 
average, $F_{ii}$, plotted on the $y$-axis, 
does not require such scaling, as it is dimensionless.

\subsection{Local density of states}

\begin{figure}
  \centerline{\psfig{figure=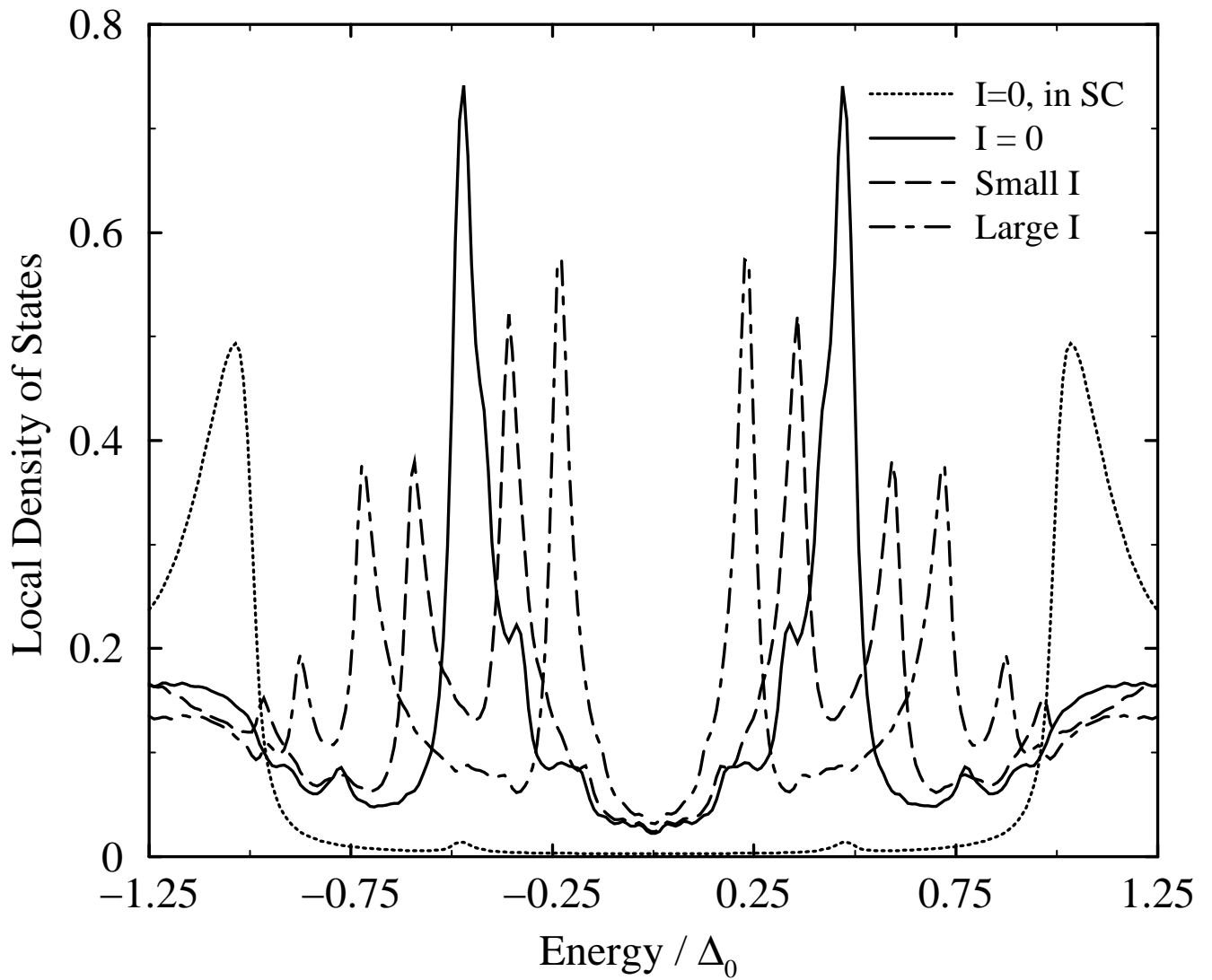,width=7.0in}}
  \caption{The local density of states at the center of a Josephson 
junction (the plane $\alpha=0$), 
with increasing current flow through the device ($N_{b}=20$). 
The solid line is for zero current, the dashed line is for small current, 
and the chain-dashed line is for large current. 
The density of 
states within the superconducting region (plane $\alpha=25$) 
at zero current flow, where 
$\Delta_{0} = 0.198$ is shown in the dotted line as a comparison. Note that 
the small bump at $E=0.5\Delta_{0}$ for the superconducting region 
arises from the self-consistency relation and the proximity of this 
superconducting plane to the barrier. }
\label{fig:ldos}
\end{figure}

\begin{figure}
  \centerline{\psfig{figure=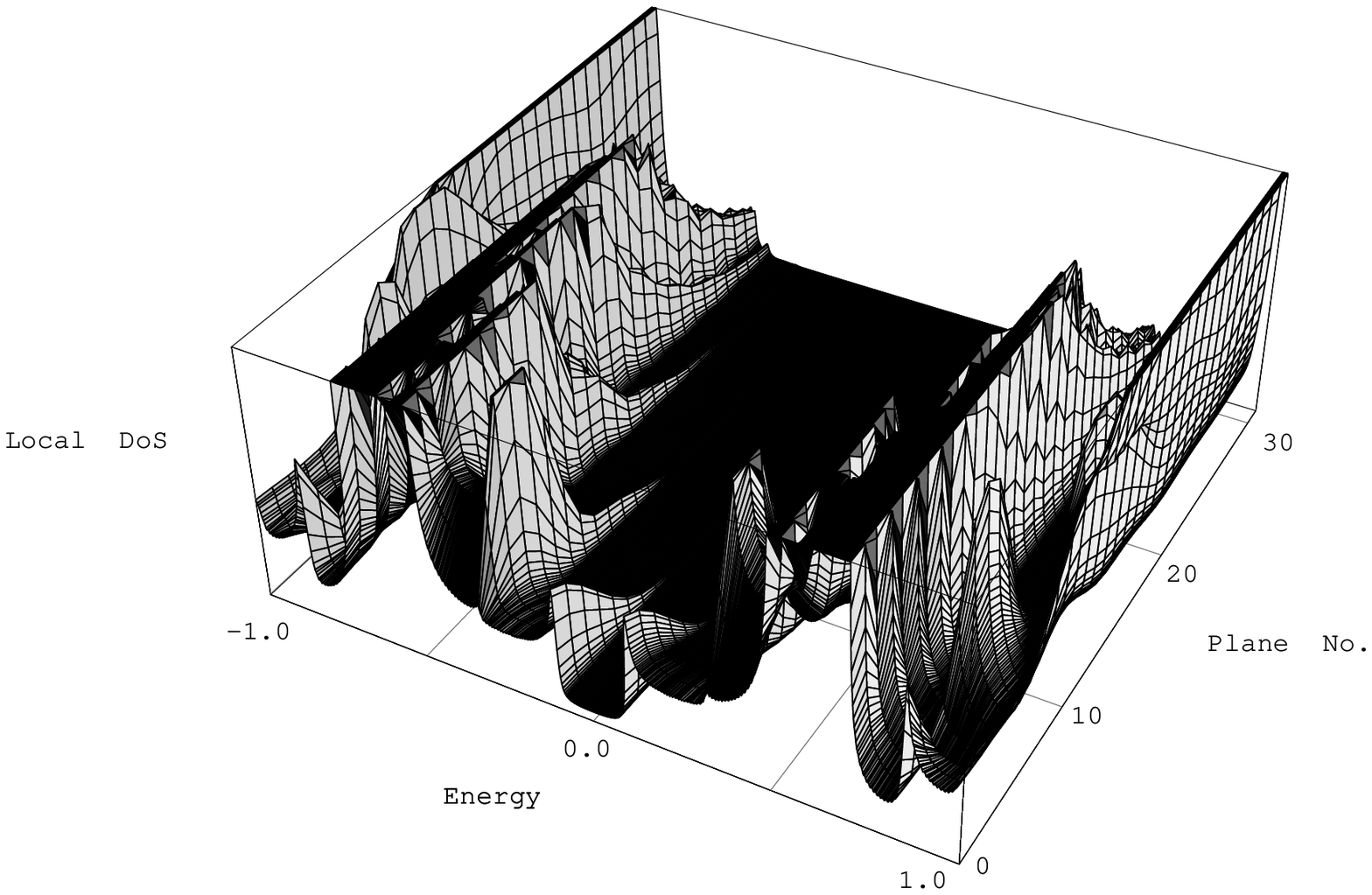,width=7.0in}}
  \caption{The evolution of the local density of states within the gap, 
as a function of position in a Josephson 
junction. The energy axis is normalized by $\Delta_{0} = 0.198$. 
Note there are 
even parity bound states, with a maximum at the center plane, and odd 
parity states with a node at that point. The self-consistent solution is 
for a barrier of width 10 planes ($N_{b}=10$). }
\label{fig:3dldos}
\end{figure}

It is interesting to observe the density of states, in particular the 
presence 
of states within the gap as shown in Fig.~\ref{fig:ldos}(a). 
The two major peaks correspond to Andreev bound states, 
which carry current, when there is a 
phase difference across the barrier region. The bound states can be seen to 
split in two, so that two inner peaks move towards each other, while two 
outer peaks move to the gap edge as the current flow increases. 
A careful examination of the curve corresponding to zero 
current reveals a great deal of structure, due to states trapped within 
the barrier region. The exact energies of the bound states depend on the 
width of the barrier, and the states arise in part from normal reflection of 
quasiparticles with non-zero momenta parallel to the 
planes\cite{Sipr,Kumm1,thomas}. 
The intraplanar momenta of the quasiparticles may lead to the states being 
unobservable as current peaks at the corresponding voltages 
in an I-V curve, where only quasiparticles travelling perpendicular to 
the planes are measured. In Fig.~\ref{fig:3dldos}, we examine the bound states 
in more detail, for a system with ten barrier planes ($N_{b}=10$). It is 
worthwhile noting that as the energy of the states within the gap approaches the 
gap edge, so they extend further away from the barrier, into the superconducting 
region [which begins after the fifth plane from the center ($\alpha > 5$)]. 
The figure also demonstrates an alternating parity between states. The 
states with energy closest to zero have even parity, as demonstrated by a maximum 
amplitude at the central planes ($\alpha=0,1$). 
The next higher energy bound states have a node 
at the central planes, so exhibit odd parity. 

\begin{figure}
  \centerline{\psfig{figure=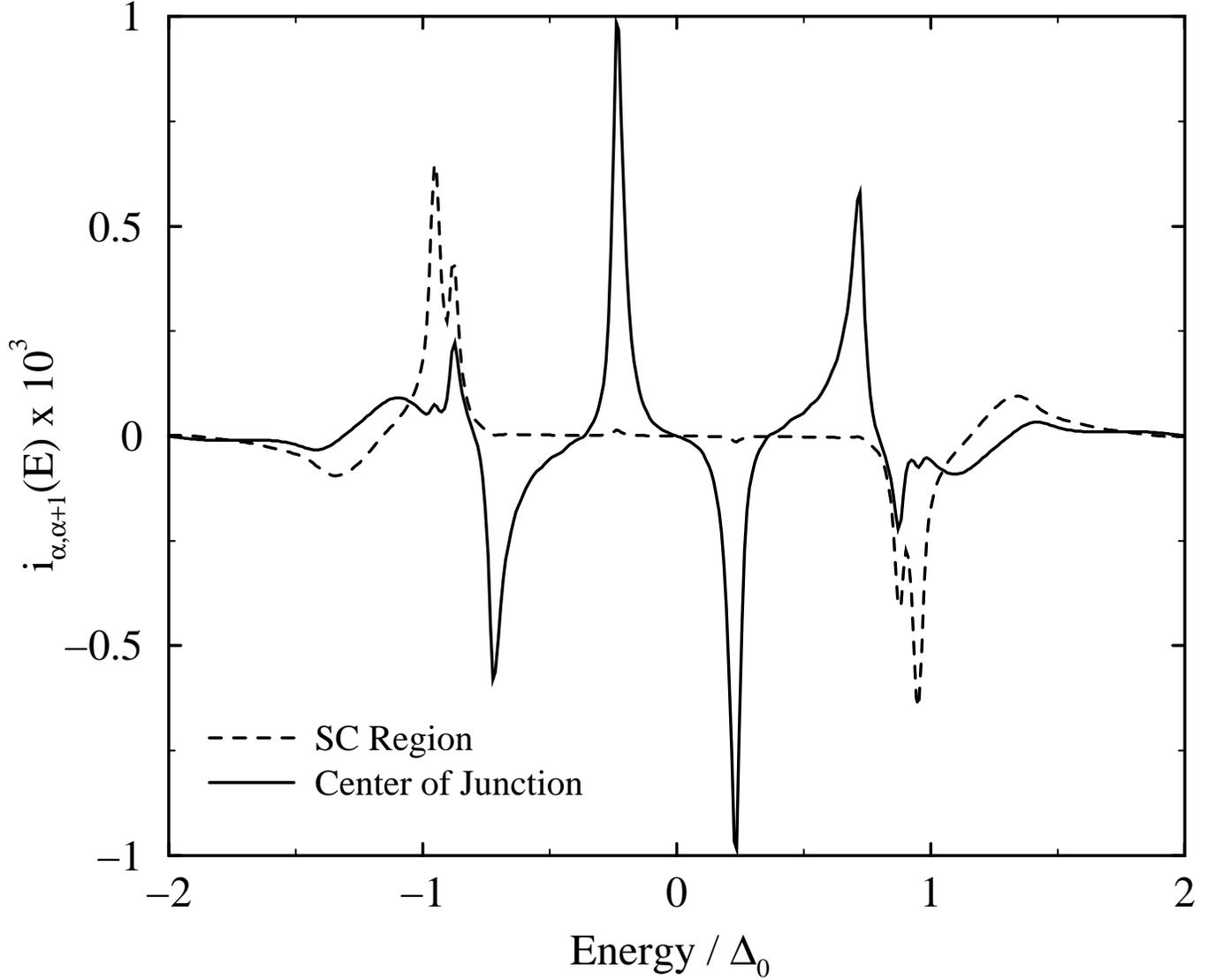,width=7.0in}}
  \caption{The local current-carrying density of states, 
$i_{\alpha, \alpha+1}(E)$ 
at the center of a Josephson junction (where $\alpha=0$) and in the 
superconducting region (where $\alpha=25$). In this junction $N_{b}=20$. }
\label{fig:jdos}
\end{figure}

We can define a current-carrying local density of states such that the total 
current, $I_{\alpha,\alpha+1}$, between two successive planes, 
$\alpha$ and $\alpha+1$, is given by:
\begin{equation}
I_{\alpha,\alpha+1} = \int i_{\alpha,\alpha+1}(E) dE.
\end{equation}
A plot of the function $i_{\alpha,\alpha+1}(E)$ between the two planes, 
$\alpha=0$ and $ \alpha+1=1$, 
at the center of the barrier is shown in Fig.~\ref{fig:jdos}. It can be 
seen that the majority of the current is carried by the Andreev bound states, 
and that the two peaks that have separated from a single Andreev peak at zero 
phase difference, carry current in opposite directions. The states at positive 
energy carry equal and opposite current to the states at negative energy, but 
their occupation is much lower for low temperatures, $T<T_{c}$. 

\subsection{Charge Depletion or Accumulation Regions}

By including an extra repulsive or attractive potential on the interfacial 
planes which connect the barrier to the superconductor, we are able to model 
in a very simple manner, some of the effects of a charge accumulation region 
or a Schottky Barrier. In our simple model there is a mirror symmetry 
for a half-filled band : a repulsive potential depletes the electron density 
in one layer, to reproduce some effects of a Schottky Barrier, while an 
attractive potential results in a charge accumulation region. The effect on the 
pairing potential and current density are the same, for equivalent potential 
strengths at half-filling. That is, if a repulsive interfacial 
potential, $V_{1}$, results 
in a reduction in electron density on a specific plane, then the attractive 
potential, $-V_{1}$, at the interface, causes an equal in magnitude 
increase of electron density at that plane, and results in a system with the 
same variation in order parameter, and equal current response.

\begin{figure}
  \centerline{\psfig{figure=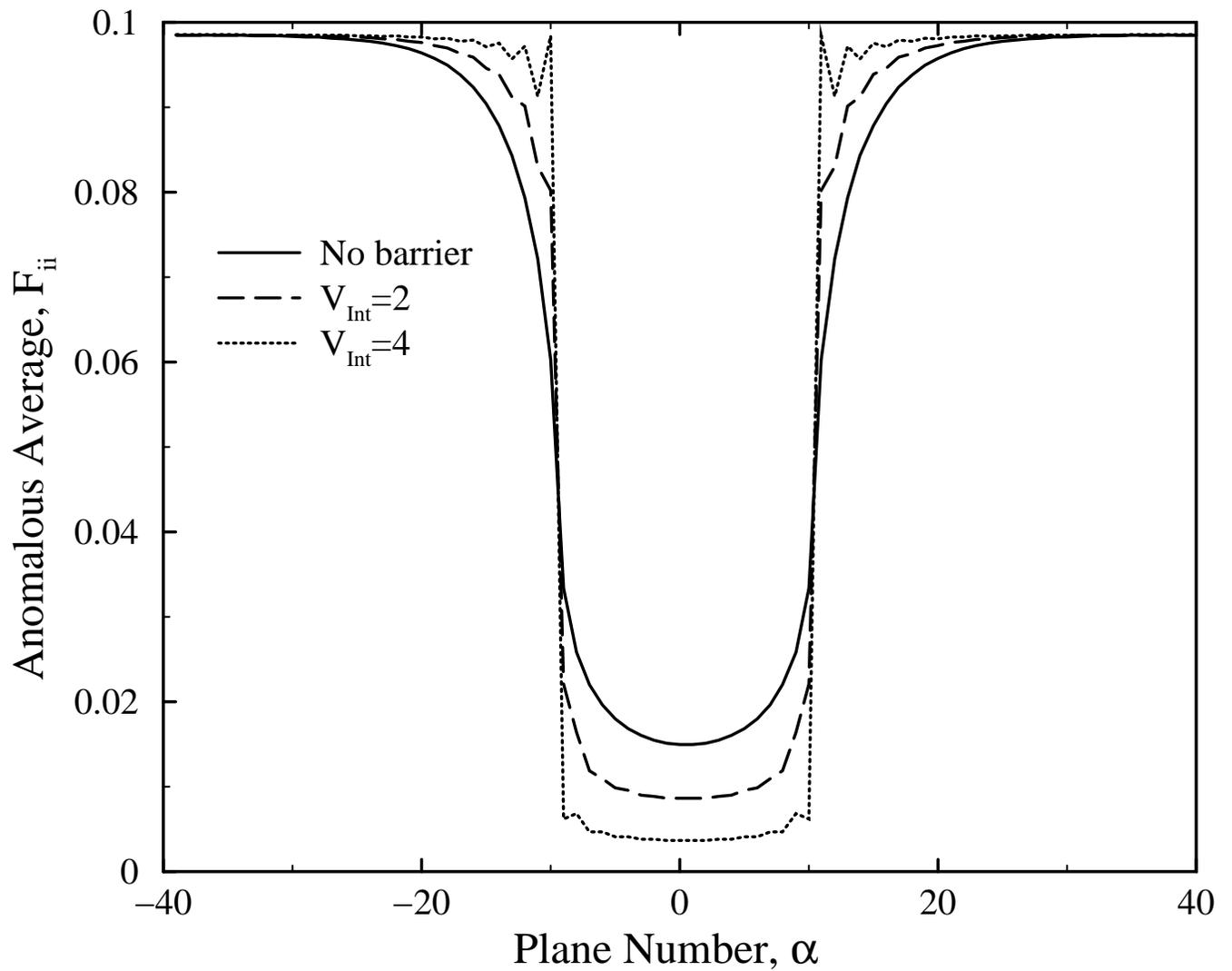,width=7.0in}}
  \caption{ The off-diagonal local Green's function, $F_{ii}$, indicating Friedel 
oscillations and a reduced proximity effect as the interfacial scattering is 
increased.}
\label{fig:mubarr1}
\end{figure}

Fig.~\ref{fig:mubarr1}(a) indicates the oscillations in the anomalous 
average resulting from a single layer barrier with $V_{0} = \pm 2t$ and 
$V_{0} = \pm 4t$. Notice that the proximity effect is reduced by the 
additional scattering potential, so the order parameter exhibits variation 
more like a step-function, with increasing barrier strength. We can think 
of the interfacial scattering as modifying the junction from $SNS$ to $SIS$ 
characteristics. 

\begin{figure}
  \centerline{\psfig{figure=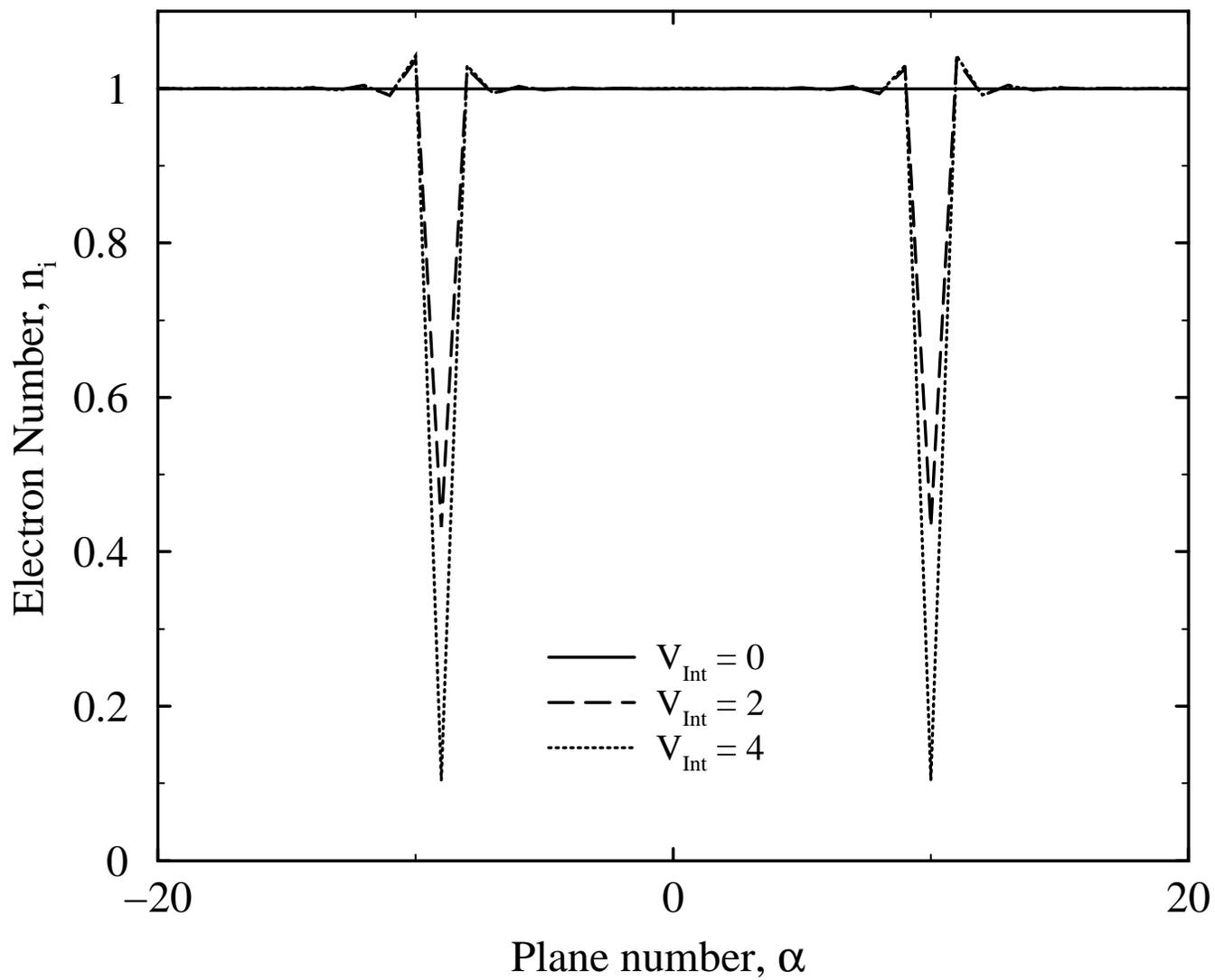,width=7.0in}}
  \caption{ Friedel oscillations in the electron density due to an 
 interfacial scattering potential.}
\label{fig:mubarr2}
\end{figure}

Fig.~\ref{fig:mubarr2} depicts the effect on the electron density 
of a positive interfacial potential, which is like a Schottky barrier, since the 
electronic charge is depleted at the interface. The charge 
density also exhibits Friedel oscillations away from the barrier. 
The transformation 
$n_{i} \mapsto 2-n_{i}$ maps results for positive potential barrier to 
negative potential, $V_{0} \mapsto -V_{0}$, and will change the system to 
that of a charge accumulation region. 

\section{Charge Impurity Scattering}

We model a barrier region with impurities by using the Falicov-Kimball model, 
as described in Section II, using 
Eq.~(\ref{eq:hhub}) with $U^{FK}<0$ and using the self-consistency 
procedure of Eqs.(\ref{eq:itera}-\ref{eq:iterc}), 
as shown in Fig.~\ref{fig:iter}. 
We carry out the calculations with an impurity concentration, 
$\rho_{imp}$ that ranges from 0.01 to 0.2. In the limit of $\rho_{imp}=0$ 
there is no scattering in the barrier, and the results correspond to 
the Hubbard model with $U_{b} = 0$. Impurities 
in the barrier region lead to an imaginary part for the frequency-dependent 
self energy and, since the lifetime of the quasiparticles at the 
Fermi surface becomes finite, to a non-Fermi liquid, characteristic of 
annealed disorder scattering. 

\subsection{Current Responses}

\begin{figure}
  \centerline{\psfig{figure=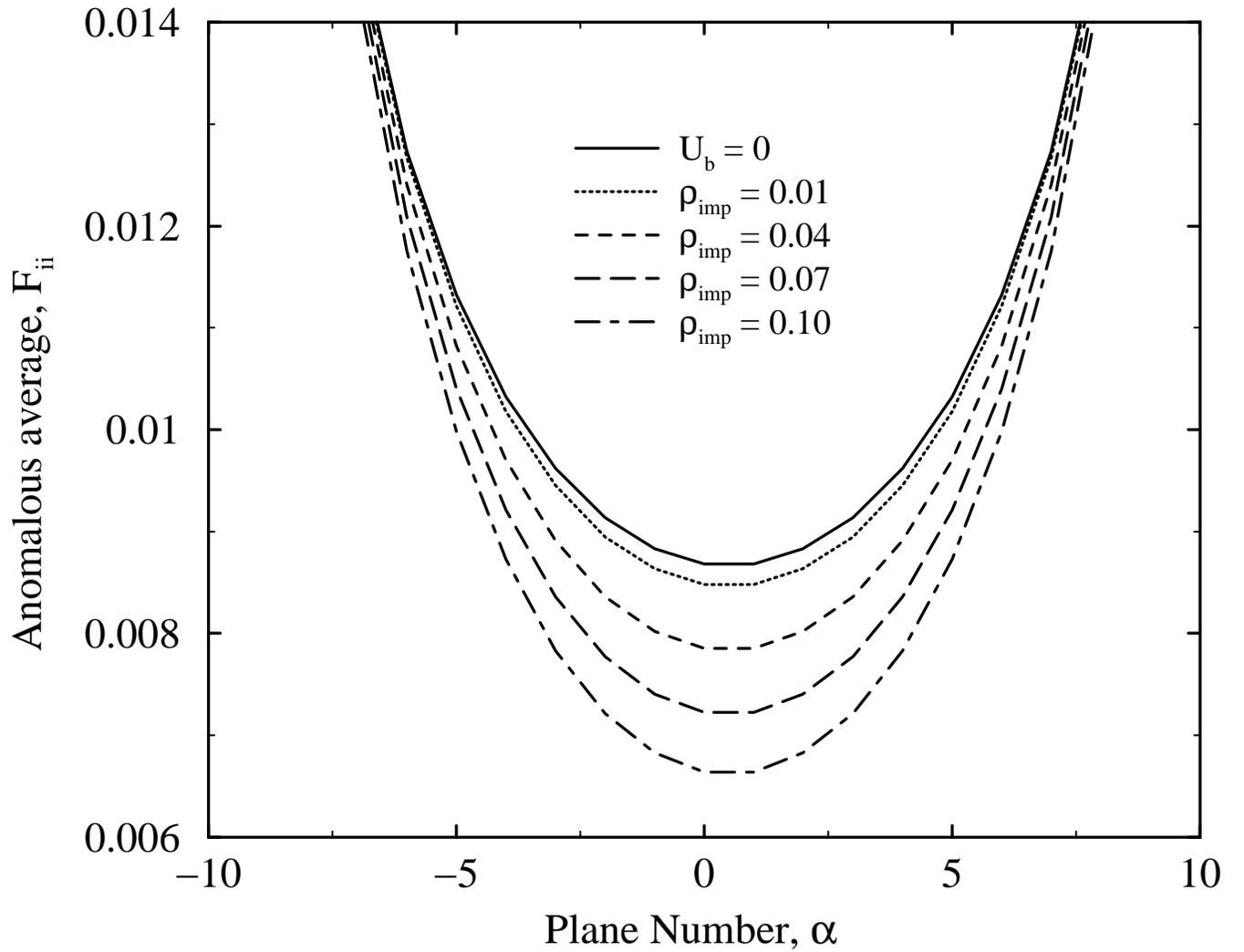,width=7.0in}}
  \caption{ The reduction of the anomalous average, 
$F_{ii}=F({\bf r}_{i},{\bf r}_{i},\tau=0^{+})$, for planes near the center of the 
barrier, with increasing impurity concentration, $\rho_{i}$, and 
$U^{FK}=-2$. }
\label{fig:fkdel}
\end{figure}

\begin{figure}
  \centerline{\psfig{figure=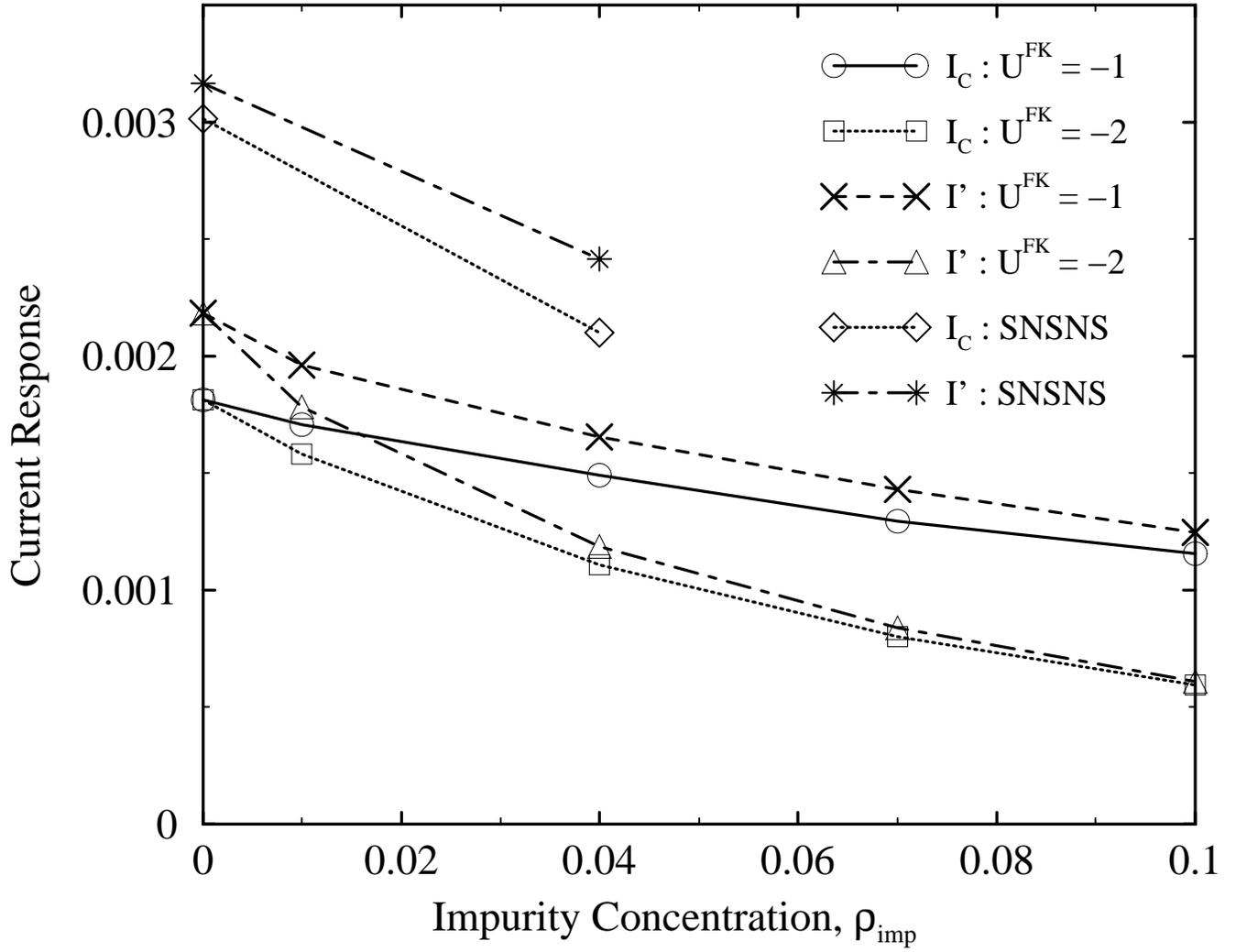,width=7.0in}}
  \caption{ The critical current, $I_{c}$, and linear response current, $I'$, 
as a function of charge impurities. Note that $I_{c}$ and $I'$ 
decrease as impurities are added to the barrier region. The greater 
scattering potential of $U^{FK}=-2$ results in greater decrease in current 
responses, compared to a scattering potential of $U_{FK}=-1$. If 
superconducting planes are inserted in the middle of the 
barrier with $U^{FK}=-2$ (SNSNS) the 
critical current is seen to increase dramatically.}
\label{fig:fkic}
\end{figure}

The addition of a small number of impurities is seen in Fig.~\ref{fig:fkdel}
to decrease the anomalous 
average in the barrier. The accompanying decrease in both critical current 
and linear response current, shown in Fig.~\ref{fig:fkic} by the 
curves labeled $U_{FK}=-2$, is more severe. 
Addition of 10\% 
impurities ($\rho_{imp}=0.10$) leads to a reduction in  both current 
responses to approximately $1/3$ of their initial values, while the 
anomalous average remains at approximately $3/4$ its original amplitude. 
Anderson's theorem, which states 
that non-magnetic impurities do not detract from the superconducting properties 
of a system only holds for a spatially homogeneous system. 
A current flow breaks 
the symmetry, and a Josephson junction is inhomogeneous in one dimension, so 
the effects we observe do not violate Anderson's theorem. General 
considerations of Green's functions in a homogeneous system show that 
increasing the imaginary part of the electronic self-energy (hence reducing 
the quasi-particle lifetime) leads to a reduction in supercurrent for a given 
phase gradient. Hence it is expected that impurities would have a more 
deleterious effect on current responses than other superconducting properties, 
which is in the spirit of Anderson's theorem.

\begin{figure}
  \centerline{\psfig{figure=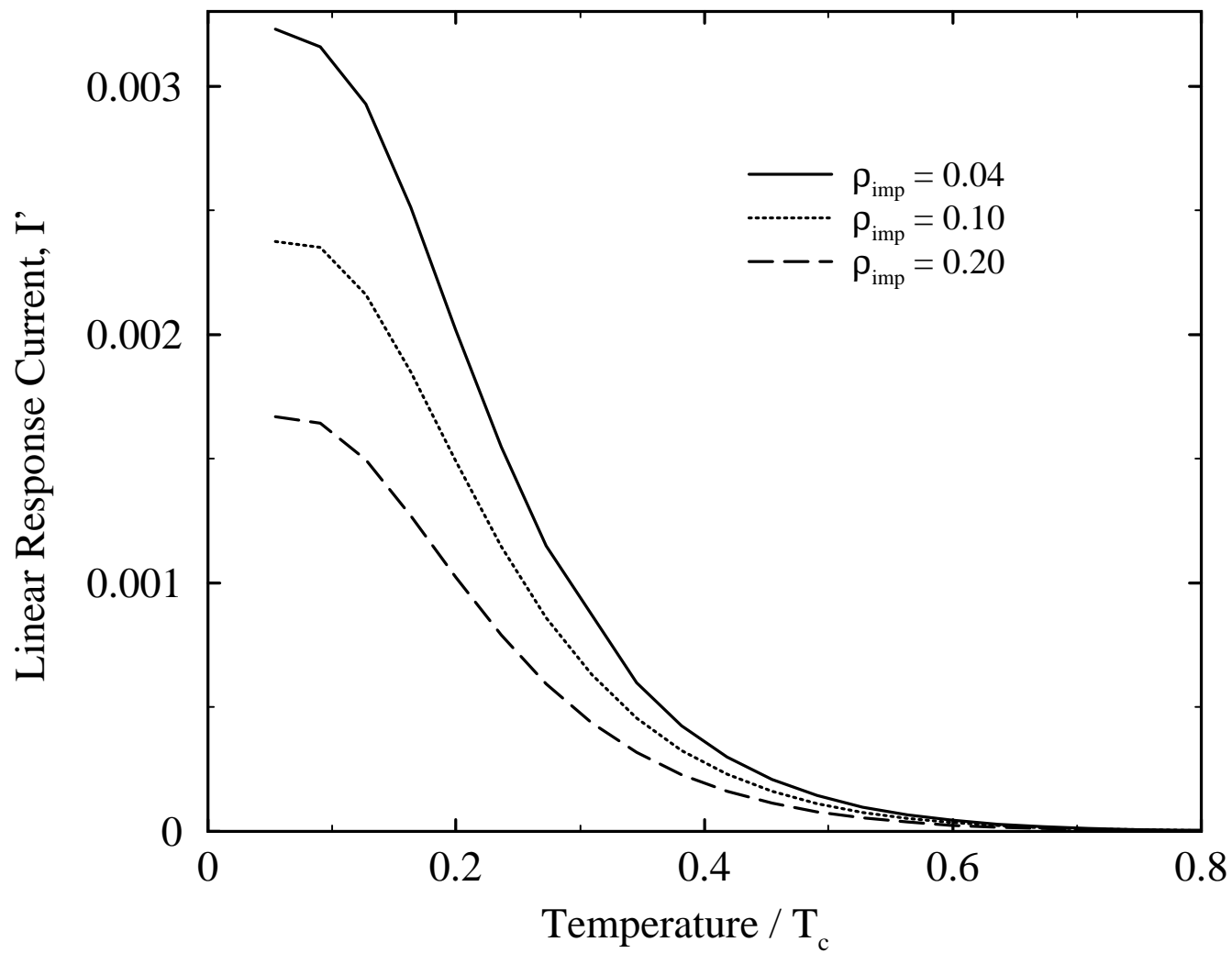,width=7.0in}}
  \caption{The linear response current, $I'$ as a function of temperature 
for increasing impurity concentrations in the barrier. }
\label{fig:ilin}
\end{figure}

\begin{figure}
  \centerline{\psfig{figure=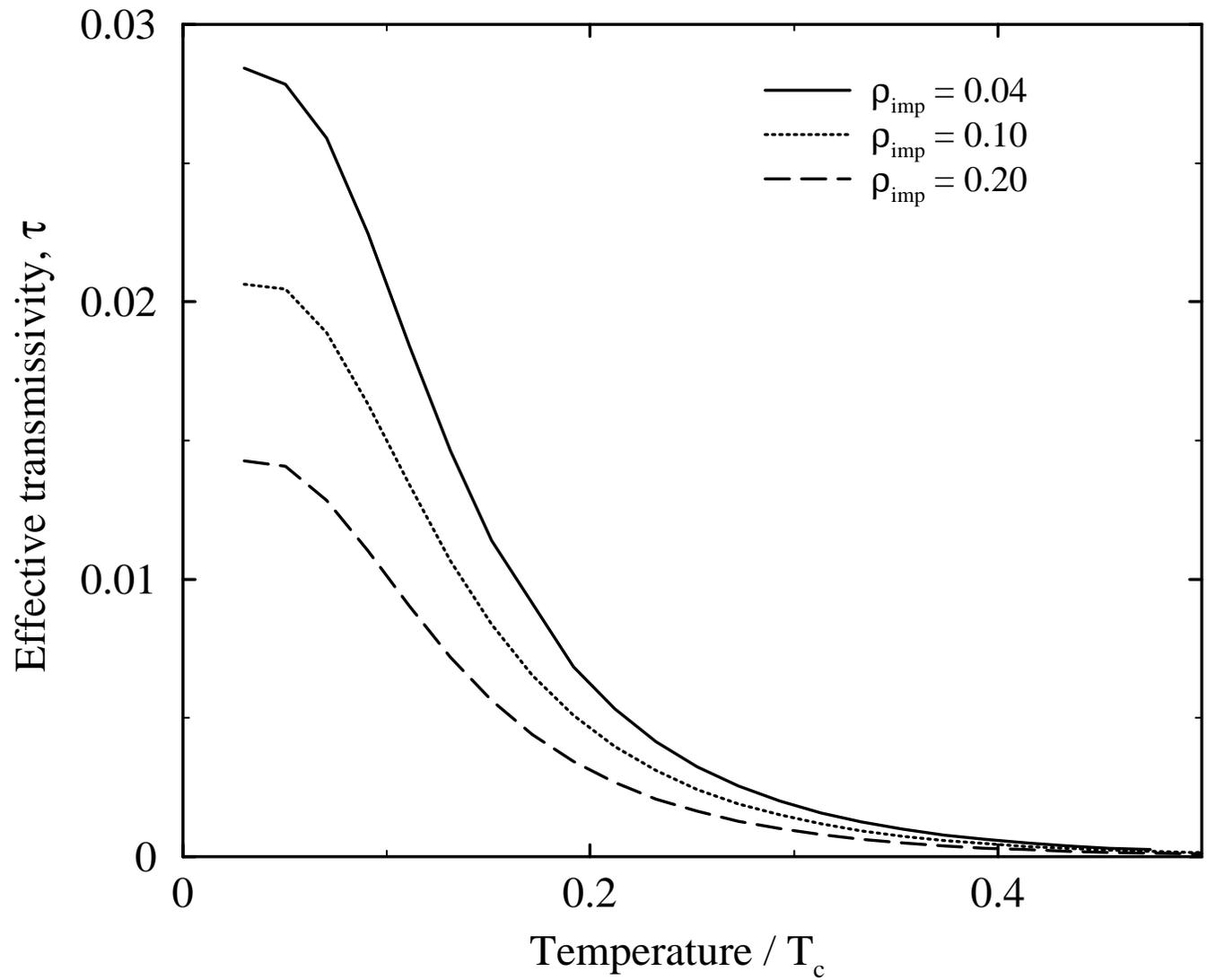,width=7.0in}}
  \caption{The effective transmissivity as a function of temperature. 
Note how the transmissivity decays with increasing temperature, 
shown here for barriers containing impurity scatterers. }
\label{fig:transmiss}
\end{figure}

We look at the linear response current, $I'$, as a function of temperature for a 
barrier with impurity scattering, in Fig.~\ref{fig:ilin}. 
The critical current remains slightly above zero until the bulk critical 
temperature, $T_{c}$. 
As suggested by Eq.~(\ref{eq:linearphase}), we can extract an 
effective transmissivity, $\tau = G_{N}.h/2e^{2}$ 
of the junction from $I'$ using: 
\begin{equation}
G_{N} = I' / \left\{
\frac{\pi \Delta_{0}(T) }{2e} 
\tanh \left( \frac{ \Delta_{0}(T) }{2k_{B} T} \right)
\right\}
\end{equation}
where $\Delta_{0}(T)$ is the value of the order parameter on the last 
superconducting plane before the barrier region. Fig.~\ref{fig:transmiss} 
depicts the results, which demonstrate that a real barrier has lower 
transmissivity with increasing temperature. This can be understood from the 
variation of the self-consistent order parameter, which is reduced to zero 
within the barrier as temperature is increased. These results are impossible 
to predict using more conventional approaches, and only arise from a microscopic 
model that includes self-consistency.

\subsection{Resistance Calculations}

We find self-consistent solutions of a system in the normal state, with no 
current flow, by setting the order parameter to zero on all planes. 
The self energy of planes outside the barrier contains only a 
constant real part, as initially we carry out the calculations within the 
Hartree-Fock approximation. Given the set of local self energies, the Green's 
functions coupling any two planes are readily found, for any momentum 
parallel to the planes. We are interested in the longitudinal components 
in the z-direction (perpendicular to the uniform planes) of the conductivity 
matrix. We define the conductivity tensor for our effectively one-dimensional 
system, from the linear current response $I_{\alpha,\alpha+1}$ across a link 
between two planes, $\alpha$ and $\alpha+1$, 
due to an electric field, $E_{\beta,\beta+1}$ across {\it all} links between 
planes $\beta$ and $\beta+1$:
\begin{equation}
\sigma_{\alpha,\beta} = 
\frac{\partial I_{\alpha,\alpha+1} }{\partial E_{\beta,\beta+1}}
\end{equation}
We find the conductivity matrix with frequency component, $\nu$, 
neglecting vertex corrections (which is valid for homogeneous systems 
in the large dimensional limit) to be: 
\begin{eqnarray}
\sigma_{\alpha,\beta}(\nu) 
& = &\frac{2}{\nu} \left(\frac{eat}{\hbar}\right)^{2} 
\int^{\infty}_{-\infty} \rho^{2D}(\varepsilon_{xy}) d\varepsilon_{xy}
\int^{\infty}_{-\infty} \frac{d\omega}{2\pi} \left\{ 
\mbox{\large Im} \left[ G_{\alpha,\beta+1}(\omega,\varepsilon_{xy}) \right]
\mbox{\large Im} \left[ G_{\beta,\alpha+1}(\omega+\nu,\varepsilon_{xy})  \right]+ 
  \right.  \nonumber  \\* 
& & 
\mbox{\large Im} \left[ G_{\alpha+1,\beta}(\omega,\varepsilon_{xy}) \right] 
\mbox{\large Im} \left[ G_{\beta+1,\alpha}(\omega+\nu,\varepsilon_{xy}) \right] - 
\mbox{\large Im} \left[ G_{\alpha,\beta}(\omega,\varepsilon_{xy}) \right] 
\mbox{\large Im} \left[ G_{\beta+1,\alpha+1}(\omega+\nu,\varepsilon_{xy}) \right] - 
 \nonumber \\*
& & \left. 
\mbox{\large Im} \left[ G_{\alpha+1,\beta+1}(\omega,\varepsilon_{xy}) \right] 
\mbox{\large Im} \left[ G_{\beta,\alpha}(\omega+\nu,\varepsilon_{xy}) \right] \right\} 
\left[ f(\omega) - f(\omega+\nu) \right]
\label{eq:freqcond}
\end{eqnarray}
where $f(\omega)$ is the Fermi-Dirac distribution function and 
$\rho^{2D}(\varepsilon_{xy})$ is the two-dimensional tight-binding 
density of states, used for the sum over momenta parallel to the 
planes. 

We are interested in the zero-frequency response, which is found from the 
appropriate limit of Eq.~(\ref{eq:freqcond}):
\begin{eqnarray}
\sigma_{\alpha,\beta}(0) & = & \frac{-1}{k_{B}T} 
\left(\frac{eat}{\hbar}\right)^{2}
\int^{\infty}_{-\infty} \rho^{2D}(\varepsilon_{xy}) d\varepsilon_{xy}
\int^{\infty}_{-\infty} \frac{d\omega}{2\pi} \nonumber \\*
& & \frac{ 
\mbox{\large Im} \left[ G_{\alpha,\beta+1}(\omega,\varepsilon_{xy}) \right]
\mbox{\large Im} \left[ G_{\beta,\alpha+1}(\omega,\varepsilon_{xy}) \right]-
\mbox{\large Im} \left[ G_{\alpha,\beta}(\omega,\varepsilon_{xy}) \right]
\mbox{\large Im} \left[ G_{\beta+1,\alpha+1}(\omega,\varepsilon_{xy})\right] }
{\cosh^{2}\left( \omega/(2k_{B}T) \right) }
\end{eqnarray}
When calculating the resistance of the junction, it is important to be aware 
that for an inhomogeneous 1D system the current flow must be uniform but 
the electric field is not. The relationship 
\begin{equation}
I_{\alpha,\alpha+1} = I_{0} = \sum_{\beta} \sigma_{\alpha,\beta} E_{\beta,\beta+1}
\end{equation}
leads to 
\begin{equation}
E_{\beta,\beta+1} = \sum_{\alpha} \left(\sigma^{-1}\right)_{\beta,\alpha} I_{0},
\end{equation}
by multiplying on the left by the inverse of the conductivity tensor. 
The voltage across the junction is the sum of the electric fields across each 
link, multiplied by the lattice spacing, $a$, so we obtain the resistance, 
\begin{equation}
R_{N} = \frac{V}{I_{0}} = 
a \sum_{\alpha,\beta}\left(\sigma^{-1}\right)_{\beta,\alpha}
\end{equation}
given by the sum of components of the inverse conductivity tensor. 

It is worthwhile pointing out, that where there is no imaginary part to the 
self energy, so there is no quasiparticle decay, the conductivity tensor 
consists of constant elements, $\sigma_{\alpha,\beta} = \sigma_{0}$. 
In such a region, 
the electric field required to produce a current flow approaches zero as the 
inverse of the system size, so the local conductivity becomes infinite. 
(That is, 
the sum of elements in one row of the conductivity tensor increases with 
system size.) 
However, the voltage drop across the region, given by the 
product of the electric field and length of perfect lead, remains constant (equal 
to $I_{0}a/\sigma_{0}$, so the 
resistance is non-zero~\cite{Butl} (equal to $a/\sigma_{0}$) while the local 
resistivity vanishes with large system size. 
In our calculations, where we neglect any lifetime 
effects of electrons outside the barrier region 
(in the Hartree-Fock approximation), there is still a contribution 
to the resistance of the junction from the `perfect' leads, but the value of 
the contribution does not depend on the length of the leads, so it can be 
thought of as a contact resistance. 

\begin{figure}
  \centerline{\psfig{figure=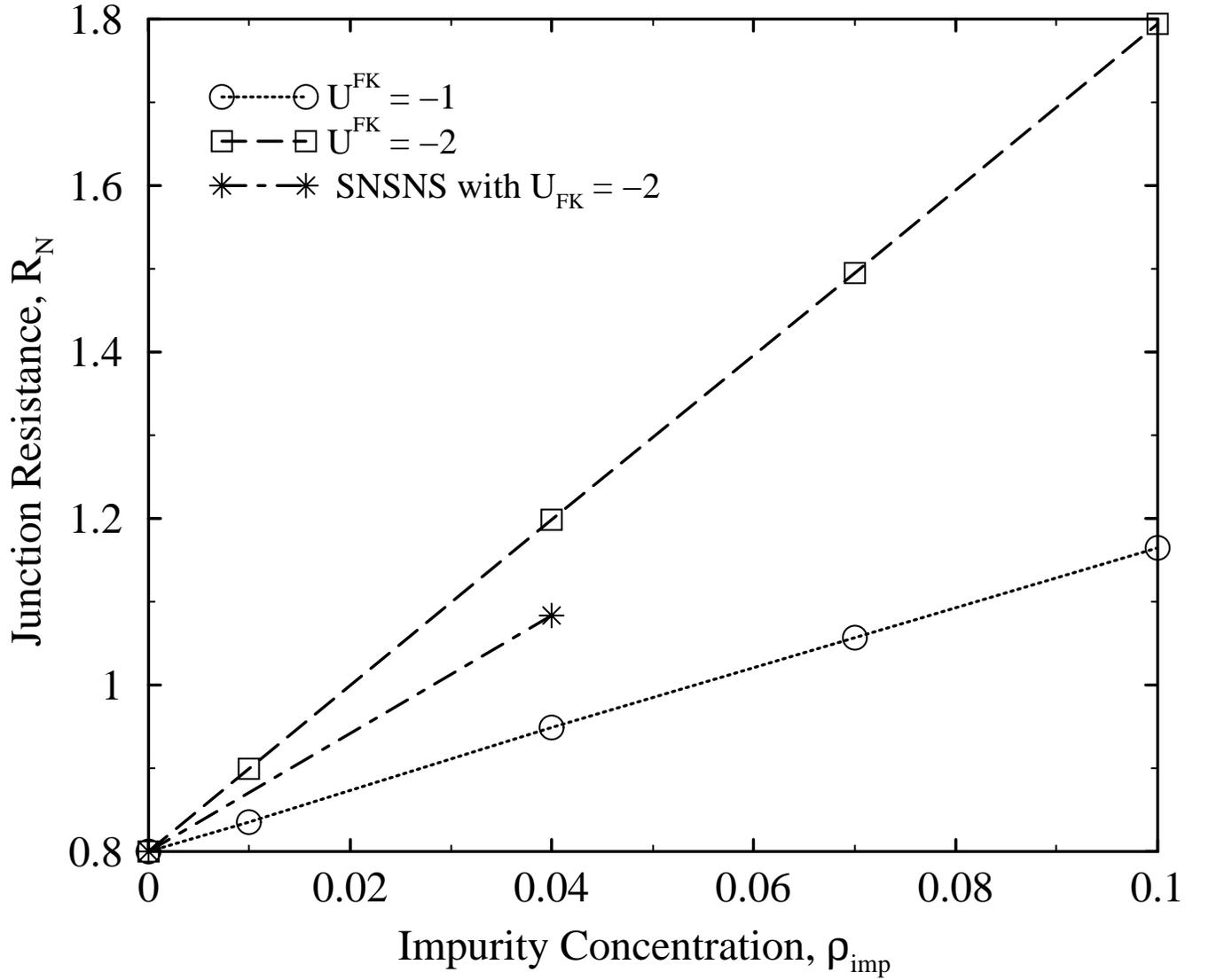,width=7.0in}}
  \caption{ The normal state resistance of a junction versus impurity 
concentration 
in the barrier. The scattering potential is greater, $U^{FK}=-2$ 
for the higher 
resistance curve compared to $U^{FK}=-1$ for the lower curve. Insertion of 
extra planes of superconducting material in the center of the barrier 
with $U^{FK}=-2$ (SNSNS)
reduces the resistance in this case.}
\label{fig:fkres}
\end{figure}

Fig.~\ref{fig:fkres} indicates the variation of junction resistance with 
impurity concentration in the barrier. The resistance increases linearly 
with number of scattering sites for the small concentrations calculated, with 
the slope increasing with the strength of scatterers. The intercept is at a 
finite resistance, which corresponds to the resistance of an infinitely long, 
perfectly conducting lead, with conductivity tensor given by 
$\sigma_{\alpha,\beta} =\sigma_{0} \approx 1.25$. 
The resistance calculated for junctions with impurity scattering within the 
barrier region, does not change when the number of 
perfectly conducting planes on either side of the barrier is increased 
from one to twenty-five. Thereafter, numerical instabilities in the matrix 
inversion process make the calculations unreliable, but we can be confident 
that the answer already arrived at is the appropriate one for the 
infinite system.

The product $I_{c}.R_{N}$ decreases with increasing concentration 
of impurities in the barrier for the examples shown. That is, the reduction 
in critical current is greater than the increase in resistance due to 
impurity scattering. A system which increases the $I_{c}.R_{N}$ product is 
found by incorporating an extra (coherent) 
superconducting region within the barrier~\cite{SNSNS}. 
Such an $SNS'NS$ or $SIS'IS$ structure has been examined, where an $SNS$ structure 
with 20 barrier planes has its central 6 barrier planes replaced with 
superconducting material, creating a barrier sandwich of 7 normal, 
6 superconducting then 7 more normal planes. Both the critical current and 
linear response current increase by a factor of greater than two, while the 
normal state resistance is only reduced by 15\% 
from its value for the SNS junction.

\section{Conclusions}

We have developed an effective method for calculating the equilibrium 
properties 
of Josephson junctions. We are able to examine the microscopic details of 
self-consistently solved 
systems through local and non-local Green's functions. The importance of 
self-consistency has previously been shown~\cite{Annet,Annet2}, 
so we go beyond 
standard `potential barrier' models of junctions to include the effects of 
spatial correlation and local fluctuations on the self-consistent potentials. 

Our results are interpreted in terms of a linear-response current, due to 
a small phase difference across the junction, as well as the critical 
current, being the maximum current that a junction can sustain. We find that 
a self-consistent microscopic determination of the potentials in the system 
for different phase differences results in current-phase behavior not 
predictable by standard fixed potential, semiclassical approaches. We study the 
effects of a number of properties of the barrier material, including its 
electron-electron interaction potential, its hopping integral and its charge 
impurity concentration. We also study interfacial scattering potentials, which 
mimic Schottky barriers and charge accumulation regions, and include barriers 
with a range of thicknesses. As well as quantifying the change in current response 
due to such modifications in the barrier, we also depict the alterations in 
the proximity effect for all cases, and charge variation in the case of 
an interfacial potential. In future work, we plan to model the charge 
redistribution at interfaces more realistically, by incorporating a non-local 
Coulomb potential, which will self-consistently determine an effective 
potential with the charge density for each plane. By such a model, we expect 
to discover if there is any significant charge redistribution as a junction 
passes through its superconducting transition, as suggested by 
Greene's group~\cite{Greene}.

We have also plotted the local density of states, to observe 
the bound states which occur at energies less than the bulk gap, 
within the barrier region. These 
states can be both current-carrying Andreev states, or states arising from 
normal reflections. Their evolution with position and phase variation is 
seen within our model, as is their contribution to the total current flow 
when a phase difference is applied. We plan in later work to show how the 
detailed structure of electronic states, apparent in these equilibrium results, 
will affect the I-V characteristics of a junction. 

We have carried out resistance calculations for junctions with impurities 
in the barrier, and found that in general the reduction in critical current 
outweighs the increase in resistance due to charge impurities. 
We suggest that a junction of the form $SNSNS$, where a thin layer of 
superconductor is placed within an normal metal barrier, can increase the 
critical current of a junction dramatically, without markedly reducing its 
resistance. We are in a position to study how more subtle effects involving 
electronic correlation close to the 
metal-insulator can affect the $I_{c}.R_{N}$ product. We plan to progress 
by making contact with specific materials, such as a Nb-InAs-Nb junction. 

Our microscopic formulation will be particularly necessary when we proceed 
to study junctions created with high-temperature superconductor materials. 
The d-wave symmetry of the order parameter, with its directional dependence, 
results in behavior that is not addressed by models utilizing a single 
transmissivity function. While an appropriate non-local version of the 
Bogoliubov-de Gennes equations can provide some insight into the properties 
of d-wave junctions~\cite{Annet,Annet2}, 
use of the DCA~\cite{DCA1,DCA2} is necessary to tackle the problem in 
any realistic manner.

\section{Acknowledgements}
We are grateful to the Office of Naval Research for funding under grant 
number N00014-99-1-0328.

\begin{figure}
  \centerline{\psfig{figure=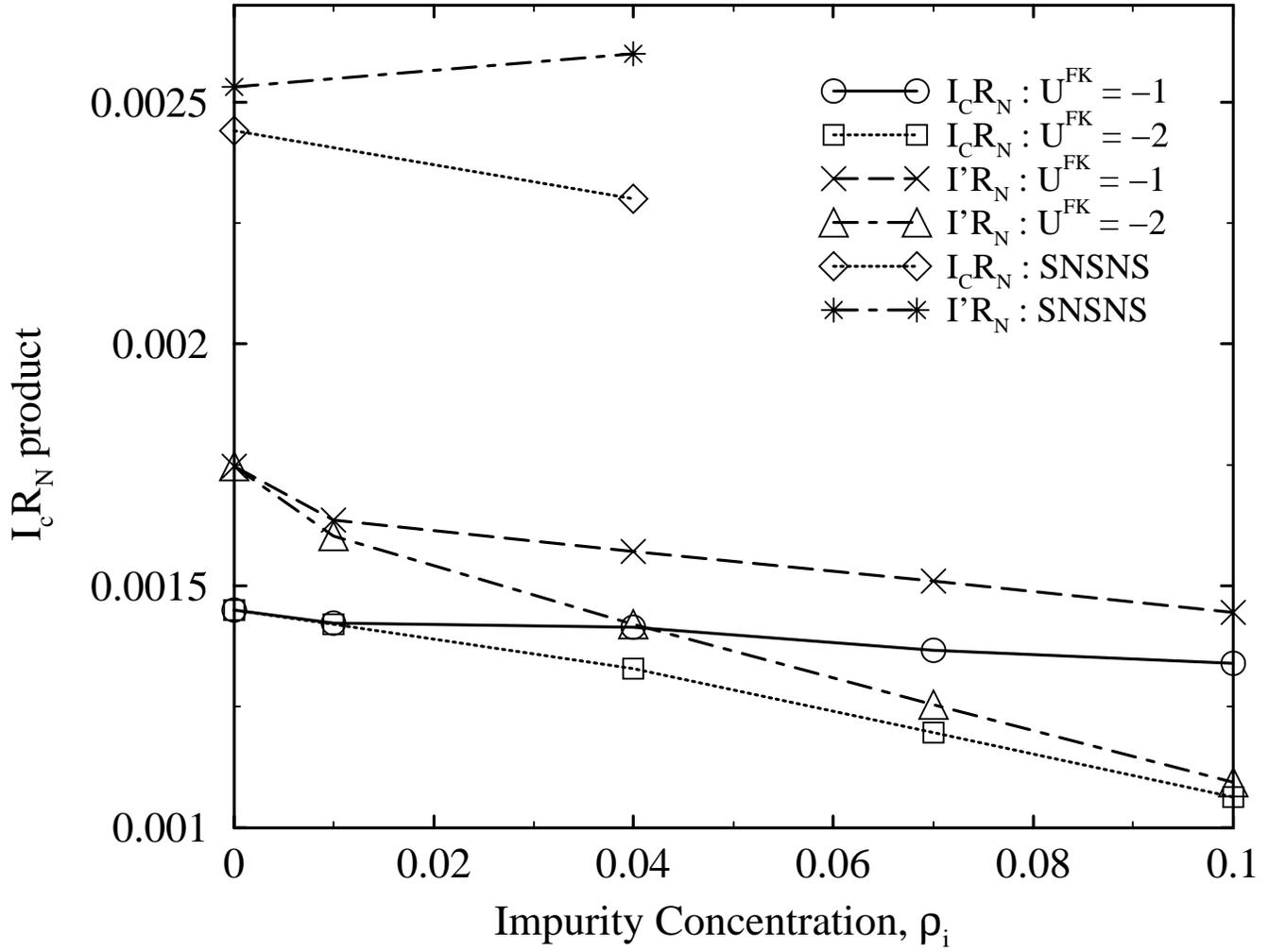,width=7.0in}}
  \caption{ The product of normal resistance of a junction  
with critical current, $I_{c}$, and linear response current, $I'$, 
for a junction with impurity scattering in the barrier. Scattering 
strengths of $U^{FK}=-1$ and $U^{FK}=-2$ are plotted. Increased figures of 
merit occur for both clean systems and for the SNSNS structures. } 
\label{fig:fkicrn}
\end{figure}

\end{document}